\documentclass[a4paper,11pt]{article}
\usepackage{amsmath,amsfonts,amssymb,times,mathptmx,graphicx,amsthm,lineno}
\usepackage{placeins}
\usepackage[T1]{fontenc}
\usepackage[latin1]{inputenc}

\title{\bf  Problems of Position Reconstruction in Silicon Microstrip Detectors
 }
\author{Gregorio Landi\thanks{Corresponding
author. Gregorio.Landi@fi.infn.it}\\
Dipartimento di Fisica e Astronomia,
Universita' di Firenze and INFN, Sezione di Firenze,\\
Largo E. Fermi 2 (Arcetri) 50125, Firenze, Italy\\
\\
{ October 20, 2021 and August 11, 2005}}
\date{ }

\begin{document}
\maketitle 

\begin{abstract}
The  algorithms for position reconstruction in silicon micro-strip
detectors are studied, and the signals of a minimum ionizing
particle are simulated. The center-of-gravity distributions of the
data events allow the fine tuning of the signal forms and of the
strip response functions. In the sensors with floating strips the
response function turns out to roughly approximate the response of
a triangular function for the size of the signal involved. The
simulations are extended to non orthogonal incidence. In these
directions, and in general for all the asymmetric signal
distributions, the standard application of the $\eta$ algorithm
introduces a systematic error. A signal reconstruction theorem
gives the way to implement the corrections of this error.
\end{abstract}

{\bf Keywords}: Center of gravity, $\eta$-algorithm, Position reconstructions, Silicon detectors.

\tableofcontents

\section{Introduction 2021}\indent

Position reconstructions for a special type of silicon micro-strip
detectors were the argument of this work.
This type of double-sided micro-strip detectors was developed for the 
ALEPH vertex detector, and installed also in the 
L3 micro-vertex detector. However, during our participation to
the L3 experiment, our attention was attracted by the
track reconstruction in the L3 central chambers 
using  the spin glass theory. 
The definition of a spin, required
the sole hit positions, thus, the properties of the
signals produced by the micro-strip detectors were 
completely neglected. 
Also the developments of refs.~\cite{landi01,landi02}
were conceived principally for
electromagnetic calorimeters, only marginally devoted
the silicon trackers. However, the availability of a
sample of test beam data of those double sided detectors
(for the PAMELA experiment)
raised our interest for an application of
the methods of ref.~\cite{landi01}. Though,
some new developments were necessary, the limitation
of the $\eta$-algorithm to symmetric systems came in 
immediate evidence. The initial constant of the 
algorithm could be easily selected only in special 
cases (i.e. for orthogonal incidence without magnetic field).
The recommendations of ref.~\cite{belau}, of an use for symmetric
systems, was completely neglected
in the implementation of the algorithm in data 
analysis of running experiments.
The warning about this point probably
originates few abandons of the $\eta$-algorithm in favor of the
simpler center of gravity. Instead, a careful study of the
simulations gave us an hint to demonstrate a first method of
correcting this systematic error in absence of magnetic field.

The efficacy of the method was tested in a dedicated
test beam ({\tt PoS(Vertex 2007) 048}) with excellent results.
Other more sophisticated methods were developed to correct
the algorithm also for moderate magnetic fields. A better
and simpler method  was
found as a by-product of an advanced method of
track reconstruction. In fact,
the observation of anomalies in the distributions
of errors in some of the following scatter-plots
(figure 11 for example) evidenced the
impossibility of an identical variance for the
hit errors
(a standard assumption of all the books on 
mathematical statistics) and
triggered a deep search for realistic models
of the error distributions of the observations.
These realistic probabilities, turn out to be different 
for each hit. As expected for the Landau distribution 
of charge released. The realistic probabilities 
allowed substantial increases of the resolution
for the track estimators. The developments, 
in the following, were essential for those
advanced tasks, supported also by theorems that impose
the account of the hit differences  for optimal 
fits ( {\tt arXiv:2103.03464} and therein references).

The two-columns printing 
of Nuclear Instruments and Methods in Physics Research
({\bf A 554} 2005 226) introduced few typing errors,
undetected by our proof revision,
absent in this version.

\section{Introduction}\indent

The aim of this work is an application of the developments of
ref.~\cite{landi01} and~\cite{landi02} to the position
reconstruction algorithms for silicon microstrip detectors. This
application turns out somewhat complex and probably too detailed
for the huge amount of data expected by future high energy
experiments. But, in some experiments, the alignment parameters of
the detector components are obtained through the position
reconstruction algorithms. It is evident the need of maximum
precision in these calibrations.

The detectors we are interested in are double sided silicon
sensors of the type used in ALEPH~\cite{aleph} and L3~\cite{L3}:
one side has all the strips connected to the electronics, the
other side has one strip each two connected to the electronics.
The unconnected strips are called "floating strips".

The first step is the simulation of the signals released in the
sensor by a minimum ionizing particle (MIP). Our simulations are
intended to be realistic enough to allow a non trivial exploration
of the position reconstruction properties for non orthogonal
incidences of a MIP. The fact that the strip charge collection
works as a powerful low pass filter assures that few parameters
are really important. To select them, we will use in reverse mode
the properties of the position reconstruction algorithms.

We proved in ref.~\cite{landi01} the strict connection of the
center of gravity (COG) algorithm with the Fourier Transform (FT)
of the signal collected. Thus, our first task will be the
calculation of the charge distribution (and its FT) that arrives
to the collecting electrodes. The signal shapes we will explore
are average signals; the fluctuations of the energy release and of
its diffusion are different in each event, but their averages are
supposed well defined. Section 2 is devoted to the determination
of the FT of the signals integrated by the strips. A parameter is
left free and extracted from the data

In the sections 3 and 4, the experimental data are used to define
the strip response function and a form factor of the signals. The
data are obtained from a test beam of the PAMELA
tracker~\cite{PAMELA}, the sensors are double sided and the
particle incidence is orthogonal. The strip property is contained
in a single function, the response function. In a rudimentary form
the strip response function integrates the energy released in its
range. The crosstalk (and the loss) modifies this picture and the
best form must be extracted from the data. For the floating strip
detectors the crosstalk is fundamental and the response function
model must be well tuned to produce faithful simulations. In this
application we will prove the effectiveness of signal
reconstruction theorems that are contained in the COG properties.

Section 5 deals with the simulation of a normal strip detector,
here the crosstalk is small, but its effect is well seen in the
data.

In section 6 the simulations are extrapolated to non orthogonal
incidence of the MIP. We will see that the asymmetry of the signal
distribution introduces a systematic error in the standard
application of the $\eta$-algorithm~\cite{belau}. A method is
defined to correct this error, the corrections are derived from a
generalization of a reconstruction theorem to asymmetric signals.

\section{The Signal Distribution}

\subsection{The Charge Diffusion}

Our task is the simulation of the signals produced by a MIP
crossing a 300 $\mu$m silicon sensor. In the absence of magnetic
field and neglecting $\delta$-ray production and multiple
scattering, the MIP releases the initial ionization along a linear
path.
\begin{figure}[h!]
\begin{center}
\includegraphics[scale=0.9]{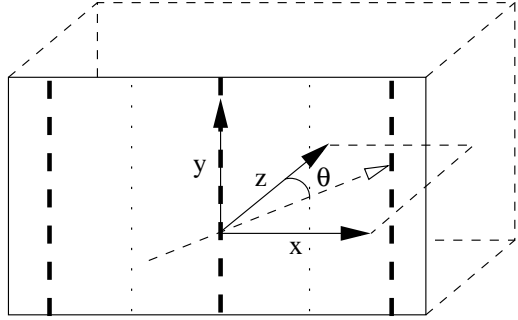}
\caption{\em Reference system on the sensor, the three dashed
thick lines are the collecting electrodes, the thin dashed arrow
is the direction of an incoming particle forming an angle $\theta$
with the $z$-axis } \label{fig:figura0}
\end{center}
\end{figure}
The electric field in the interior of the detector drifts the
charges toward the corresponding electrode. During the drift, the
charge diffusion modifies the form of the initial ionization.

The initial charge distribution (of both signs) is expressed as in
ref.~\cite{landi01}:
\begin{equation}\label{eq:110}
\varphi_0(\mathbf{r,r_0,L})=\int_0^1\mathrm{d}\lambda\delta(\mathbf{r}-
\mathbf{r}_0-\mathbf{L}\lambda)
\end{equation}
where $\delta(\mathbf{r})=\delta(x)\delta(y)\delta(z)$.
$\varphi_0(\mathbf{r,r_0,L})$ is a function of $\mathbf{r}$
defined on $\mathbb{R}^3$, it describes a uniform charge
distribution along the vector $\mathbf{L}$ and  impact point
$\mathbf{r}_0$. The impact point is on the detector surface and
the ionization path $\mathbf{L}\equiv\{L_x,L_y,L_z\}$ is contained
in the detector interior. If $L_0$ is the detector thickness, we
have $L_z=L_0$.

The diffusion process transforms an initial Dirac-$\delta$ charge
distribution in a gaussian of half width proportional to the drift
time. The final charge distribution, arriving to the collecting
electrodes, has the form:
\begin{equation}\label{eq:116}
\varphi(\mathbf{r}_0,\mathbf{r}_f,\mathbf{L})=
 \int_{\mathbf{R}^3}\mathrm{d}^3\mathbf{r}'
\varphi_0(\mathbf{\mathbf{r}',\mathbf{r}_0,\mathbf{L}})g(\mathbf{r}'-\mathbf{r}_f)
\end{equation}
where $\mathbf{r}_f=\{\mathbf{x},z_f\}$ and
$\mathbf{x}\equiv\{x,y\}$, $\mathbf{x}$ is defined on
$\mathbb{R}^2$ and $z_f$ is the plane of the collecting electrode,
$g(\mathbf{r})$ accounts the diffusion process.  The reference
system is illustrated in figure~\ref{fig:figura0}, it has the
origin in the plane of the collecting electrodes, the $z$-axis is
perpendicular to that plane and directed toward the sensor
interior. The $x$-axis is perpendicular to the strip direction.
For both charge we will put the impact point at $z=0$ and
$z_f=z_0=0$ (in one case the impact point is in reality an
outgoing point).

 The diffusion gives to $g(\mathbf{r})$ the
form:
\begin{equation}\label{eq:117}
g(\mathbf{x},t)=\frac{1}{(\sqrt{2\pi}\sigma(t))}\exp[-\frac{\mathbf{x}^2}{2\sigma(t)^2}]\,.
\end{equation}
$\sigma(t)$ is a function of the diffusion process and depends
from the time $t$ spent by the charges in their drift from the
production point to $z=0$:
\begin{equation}\label{eq:120}
  \sigma(t)=\sqrt{\mu D t}
\end{equation}
where $D$ is proportional  to the absolute temperature. The time
$t$ is connected to the electric field in the semiconductor
interior and to the charge mobility. To simplify we will take a
constant electric field $E$ in $z$ direction, and its effective
value will be extracted from the data. For $t(z)$ and $\sigma$ we
have ($z=\lambda L_0$):
\begin{equation}\label{eq:126}
  t(\lambda)=\frac{\lambda L_0}{\mu E}\ \ \ \ \
  \sigma^2=\frac{DL_0\lambda}{E}=\alpha \lambda\ \ \ \
  \alpha=\frac{DL_0}{E}
\end{equation}
More realistic field configuration could be used (with an increase
of the unknowns), but this refinement is irrelevant for the
simulation.

From equations~\ref{eq:116} and~\ref{eq:117}, and recalling our
position $z_0=z_f$,
$\varphi(\mathbf{r}_0,\mathbf{r}_f,\mathbf{L})$ does not depend on
$z_0$ and $z_f$, and it becomes:
\begin{equation}\label{eq:127}
\varphi(\mathbf{x}_0,\mathbf{x},\mathbf{L})=\int_0^1\mathrm{d}\lambda
\frac{1}{(\sqrt{2\pi \alpha\lambda})}
\exp[-\frac{(\mathbf{x}-\mathbf{x}_0-\mathbf{L}_x\lambda)^2}{2\alpha\lambda}]
\end{equation}
where we pose $\mathbf{L}_x\equiv\{L_x,L_y\}$ .
\subsection{The Fourier Transform}
The equations of ref.~\cite{landi01,landi02} are expressed with
the FT of the charge distribution respect to $\{\mathbf{x} \}$.
Defining $\Phi(\mathbf{x}_0,\mathbf{L}_x,\boldsymbol\omega) $ the
FT of $\varphi(\mathbf{x}_0,\mathbf{x},\mathbf{L}) $, its
expression is:
\begin{equation}\label{eq:140}
\begin{aligned}
&\Phi(\mathbf{x}_0,\mathbf{L}_x,\boldsymbol\omega)=\int_{\mathbb{R}^2}\frac{\mathrm{d}\mathbf{x}}{(2\pi)^2}
  \,\rho_f(\mathbf{x}_0,\mathbf{L}_x,\mathbf{x})\,\mathrm{e}^{-i\boldsymbol\omega\cdot\mathbf{x}}\\
&=\int_0^1\mathrm{d}\lambda\int_{\mathbb{R}^2}\mathrm{e}^{-i\boldsymbol\omega\cdot\mathbf{x}}
\frac{1}{\sqrt{2\pi\alpha\lambda}}\exp{[-\frac{(\mathbf{x}-\mathbf{x}_0-\mathbf{L}_x\lambda)^2}{2\alpha\lambda}]}\mathrm{d}\mathbf{x}\\
&=\int_0^1\mathrm{d}\lambda\,\mathrm{e}^{-i(\mathbf{x}_0+\mathbf{L}_x\lambda)\cdot\boldsymbol\omega}\,
\mathrm{e}^{-\alpha\lambda\omega^2/2}
\end{aligned}
\end{equation}
where we pose $\boldsymbol\omega=\{\omega_x,\omega_y\}$. The
integration on $\lambda$ gives:
\begin{equation}\label{eq:160}
\Phi(\mathbf{x}_0,\mathbf{L}_x,\boldsymbol\omega)=
\mathrm{e}^{-i\mathbf{x}_0\cdot\boldsymbol\omega}\frac{1-\mathrm{e}^{-i\mathbf{L_x}\cdot\boldsymbol\omega}
\mathrm{e}^{-\alpha\,\omega^2/2}}{i\mathbf{L_x}\cdot\boldsymbol\omega+\alpha\,\omega^2/2}
\end{equation}
The  parameter $\alpha$ collects important properties of the
detector; temperature, thickness and the effective constant field
$E$. The  values of $\alpha$ for the two sensor sides will be
extracted from the data.

The equation~(\ref{eq:160}) allows asymmetric forms for the signal
distributions easy to manage even for pixel detectors.

In our equations
$\Phi(\mathbf{x}_0,\mathbf{L}_x,\boldsymbol\omega)$ has the COG in
the origin of the reference system, this gives the relation:
\[
\lim_{\boldsymbol\omega\rightarrow
0}\frac{\partial\Phi(\mathbf{x}_0,\mathbf{L}_x,\boldsymbol\omega)}{\partial\boldsymbol\omega}
=-\frac{1}{2}i(\mathbf{L_x}+2\mathbf{x}_0)=0.
\]
In the plane $z=0$, the impact point $\mathbf{x}_0$ is related to
the to the initial ionization segment $\mathbf{L_x}$ by:
\[
\mathbf{x}_0=-\frac{\mathbf{L_x}}{2}
\]
and it has the same expression as in the absence of diffusion
($\alpha=0$). We will need impact points uniformly distributed on
a strip and, for the above equation, even their COG's. The FT
shift theorem~\cite{libroFT} allows to move the COG where needed.

In the following, we will limit our study to inclined tracks with
$L_x\neq 0$ and $L_y=0$ .

\begin{figure}[h!]
\begin{center}
\includegraphics[scale=1]{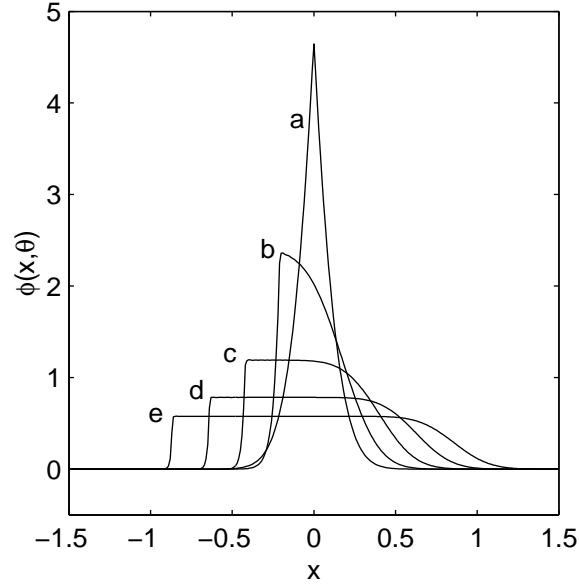}
\caption{\em Form of the signal distribution  at various incidence
angles. The curves $a$, $b$, $c$, $d$ ed $e$ are respectively for
$\theta=0^\circ$, $\theta=5^\circ$, $\theta=10^\circ$,
$\theta=15^\circ$ and $\theta=20^\circ$ } \label{fig:figura1*36}
\end{center}
\end{figure}

The equation (\ref{eq:160}) can be rearranged in:
\begin{equation}\label{eq:180}
\Phi(\mathbf{L_x},\boldsymbol\omega)=
\mathrm{e}^{-\alpha\,\omega^2/4}\frac{\sin(\frac{1}{2}\mathbf{L_x}\cdot\boldsymbol\omega-i\frac{1}{4}\alpha\omega^2)}
{(\frac{1}{2}\mathbf{L_x}\cdot\boldsymbol\omega-i\frac{1}{4}\alpha\omega^2)}.
\end{equation}
This form is an evident generalization of a track energy
introduced in ref.~\cite{landi01}.  A microstrip sensor integrates
on $y$ the signal distribution $\varphi(\mathbf{x},\mathbf{L_x})$,
and in equation~\ref{eq:180} we have $\omega_y=0$. The surviving
$\omega_x$ variable will be indicated with $\omega$, and
$\varphi(x)$ the integrated signal distribution.
Figure~\ref{fig:figura1*36} illustrates $\varphi(x)$ at various
incidence angles ($L_x=\tan(\theta)$). The COG of each
$\varphi(x,\theta)$ is in the origin $x=0$.
\subsection{Expressions for the COG and Energy}
The asymmetric signal distributions require a slight
generalization of the equations in ref.~\cite{landi01}, where
symmetric signals were assumed. The equations of the
COG-algorithms with two and three strips will be considered. As in
ref.~\cite{landi01} we indicate with
$\mathrm{S}_{\varepsilon}(\omega)$ the FT of the sampled
convolution of the signal (transformed in a periodic signal with a
period T$\gg\tau$) with the strip response. The derivative of
$\mathrm{S}_{\varepsilon}(\omega)$ respect to $\omega$ (normalized
with $\mathrm{S}_{\varepsilon}(0)$) is the COG position of the
signals collected by the strips. The parameter $\varepsilon$ is
the COG position of the initial signal distribution, and the shift
theorem of FT introduces its functional dependence in the
equations:
\begin{equation}\label{eq:200}
  \mathrm{S}_{\varepsilon}(\omega)=\sum_{n=-\infty}^{+\infty}\Phi_{n}^{p}\
\mathrm{P}(\frac{2n\pi}{T})\exp(\frac{-i2n\pi\varepsilon}{T})\mathrm{H}(\omega-\frac{2n\pi}{T})\,
.
\end{equation}
In equation~\ref{eq:200}, $\,$ $\mathrm{P}(2n\pi/T)$ is the FT of
the strip response function p$(x)$, and it will be determined in
the following. $\Phi_n^p$ is equation~\ref{eq:180} for
$\omega=2\pi n/T$ as imposed by periodicity:
\begin{equation}\label{eq:201}
  \Phi_n^p=\frac{1}{T}\Phi(L_x,\frac{2\pi
  n}{T})\, .
\end{equation}
$\mathrm{H}(\omega)$ is a function of the strip number used in the
algorithm. In the case of two strips it is ($\tau$ is the strip
dimension) :
\begin{align*}
&\mathrm{H}(\omega)=1+\mathrm{e}^{(-i\omega\tau)}\qquad
\varepsilon>0 \\
&\mathrm{H}(\omega)=1+\mathrm{e}^{(+i\omega\tau)}\qquad\varepsilon<0\\
\end{align*}
and for three strips:
\begin{align*}
&\mathrm{H}(\omega)=1+2\cos(\omega\tau)\qquad\\
\end{align*}
Due to the normalization of $\varphi(x)$ and p$(x)$,
$\mathrm{S}_{\varepsilon}(0)$ is the signal collection efficiency
of the algorithm. The efficiency for the three-strip algorithm is
given by:
\begin{equation}\label{eq:240}
\mathrm{S}_{\varepsilon}(0)=\frac{3\
\mathrm{P}(0)}{T}+2\sum_{n=1}^{+\infty}\
\mathrm{P}(\frac{2n\pi}{T})\big[ 1+2\cos(\frac{2n\pi\tau}{T})\big]
\mathrm{Real}[\Phi_{n}^{p}\exp(-i\frac{2n\pi\varepsilon}{T})]
\end{equation}
and the three-strip COG is:
\begin{equation}\label{eq:241}
x_{g3}=\frac{-4\tau}{\mathrm{S}_{\varepsilon}(0)}\sum_{n=1}^{+\infty}\
\mathrm{P}(\frac{2n\pi}{T})\sin(\frac{2n\pi\tau}{T})
\,\mathrm{Imag}[\Phi_{n}^{p}\exp(-i\frac{2n\pi\varepsilon}{T})]
\end{equation}
If in equation~\ref{eq:200} we substitute $H(\omega)=\exp(-i\omega
l \tau)$ of a single strip at $l\tau$, the expression of
$\mathrm{S}_{\varepsilon}(\omega)$ for $\omega\rightarrow 0$ gives
the fraction of the energy collected by the $l$-th strip:
\begin{equation}\label{eq:241a}
  E_\varepsilon^l=\sum_{n=-\infty}^{+\infty}\Phi_n^p\, P(\frac{2\pi n}{T})
  \exp\big[-i\frac{2\pi n}{T}(\varepsilon-l\tau)\big].
\end{equation}
Equation~\ref{eq:241a} is another way to write the convolution of
$\varphi(x-\varepsilon)$ with $p(x)$ for a strip. The periodicity
imposed to $\varphi(x)$ allows to write the convolution as Fourier
Series (FS) with all the benefits to use a fast converging
discrete sum in place of an integral. So, for the three strip COG
$x_{g3}$, we can use equation~\ref{eq:241} or use
$x_{g3}=(E_\varepsilon^1-E_\varepsilon^{-1})/(E_\varepsilon^1+E_\varepsilon^0+E_\varepsilon^{-1})$.
The last one will be preferred to be consistent with the
simulations.
\subsection{The Reconstruction of $\varphi(x)$}
Equation \ref{eq:241a} requires further information that we have
to extract from the data; a reasonable form of the strip response
function $p(x)$ (with FT $\mathrm{P}(2n\pi/T)$) and the best value
of $\alpha$ of equation~\ref{eq:180}. These sensor properties are
bound together and we must fix  their better combination with
successive tests. We have some relations to use for this task. The
relation, proved in ref~\cite{landi01}, of
$d\,x_g(\varepsilon)/d\,\varepsilon$ with $\varphi(x)$ can be used
as a starting point. In that case the response function was an
interval function ($\Pi(x)=1$ for $|x|<1/2$ and zero elsewhere).
For $d\,x_g(\varepsilon)/d\,\varepsilon$ we have ($\Phi(\omega)$
FT of $\varphi(x)$) :
\[
\frac{dx_{g}}{d\varepsilon}=1+\sum_{k=1
}^{+\infty}(-1)^{k}\mathrm{real}\Big[\exp
\big(\frac{2k\pi\varepsilon}{\tau}\big)\Phi(-\frac{2k\pi}{\tau})\Big]
\]
which for Poisson identity~\cite{libroFT2} can be reassembled into
(with $\Phi(0)=1$):
\begin{equation}\label{eq:242}
\frac{dx_{g}}{d\varepsilon}=\sum_{k=-\infty}^{+\infty}\exp\big[i\frac{2k\pi}{\tau}
(\epsilon-\frac{\tau}{2})\big]\ \Phi(-\frac{2k\pi}{\tau})=
\tau\sum_{L=-\infty}^{+\infty}\varphi(-\varepsilon+\frac{\tau}{2}-L\tau).
\end{equation}
If the support of $\varphi(x)$ is $\leq\tau$ , the multiplication
of equation~\ref{eq:242} by $\Pi(-x/\tau+1/2)$ reconstructs
exactly the function $\varphi(-x+ \tau/2)$. It is easy to extract
from the reconstruction theorem~\ref{eq:242} a relation of peaks
in the COG probability distribution and discontinuities in the
response function. In regular detector and a $\varphi(x)$ for
$\theta=0$, one has a peak in the COG probability distribution and
a discontinuity in $p(x)$ at the strip border. In a detector with
floating strip the COG probability has three peaks, and we can
guess that other discontinuities are present in the response
function.
\section{The Response Function}
\subsection{Crosstalk}\label{sec:crosstalk}
The set of candidate response function $p(x)$ will be selected
among the ones defined in~\cite{landi01} as uniform crosstalk,
i.e. response function with its influence extending outside the
range of a strip and conserving the energy (in the case of an
infinite sampling). A sufficient condition for the uniform
crosstalk  is to have arbitrary functions (even Dirac
$\delta$-function) convoluted with an interval function of size
$\tau$. In the following we will work with $\tau=1$.

This $p(x)$ is suitable to simulate the effect of a floating
strip:

\begin{equation}\label{eq:245}
\begin{aligned}
p(x)=\int_{-\infty}^{+\infty}\Pi(x-x')\big\{&a_1\;[\delta
 (x'-1/4)
 +\;\delta(x'+1/4)]\\
&+a_2[\;\delta(x'-1/2)+\;\delta(x'+1/2)]\big\}
\end{aligned}
\end{equation}
\begin{figure}[h!]
\begin{center}
\includegraphics[scale=0.9]{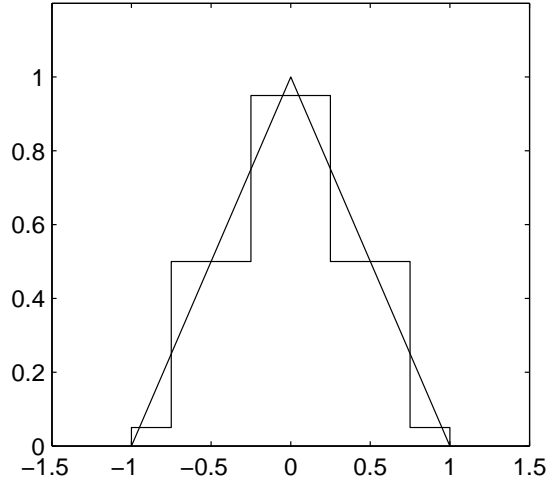}
\caption{\em Response function of equation~\ref{eq:245} with
$a_1=0.45$ and $a_2=0.05$, a type of uniform crosstalk that
reproduces the lateral peaks around $x_g=\pm 1/2$. The triangular
response function (with base 2) is reported for comparison}
\label{fig:figura1}
\end{center}
\end{figure}
It is easy to show that the $p(x)$  of figure~\ref{fig:figura1}
produces the three typical peaks of the $\eta$
distribution~\cite{belau} ($\eta=E_L/(E_L+E_R)$ where $E_L$ and
$E_R$ are the left and right strips of the two largest signal
couples) of a floating strip sensor. In fact,
figure~\ref{fig:figura1} can be read as the plot of the possible
energies collected by the strips in the case of a Dirac $\delta$
signal. The ordinate of each strip center in the array are just
the energies collected by an event. Given the strip pitch
$\tau=1$, the sample of events is generated by moving the strip
array to cover a strip interval. Two possible combination of
energy are allowed: equal energies or 0.9 an 0.05. The ratio of
one of this pair of energies divided by the sum of the two, gives
three $\delta$-like peaks at the positions 0.054, 0.5, 0.94 and
relative intensity (1 2 1). A smearing of the response function
with a realistic signal distribution and further fine tuning is
able to reproduce the experimental distributions.

The results of an experiment with an infrared laser on a double
sided silicon sensor~\cite{infrared_laser} show an energy
collection similar to figure~\ref{fig:figura1}. In their figure 6
a step-like plot is easily recognized. The due differences must be
kept in mind; the laser beam is not a Dirac $\delta$-function, the
strip metallization is opaque to the laser beam.

The response function of equation~\ref{eq:245} with $a_1=0.5$ and
$a_2=0$ and $\Phi(\omega)$ FT of $\varphi(x)$ gives for the COG
(infinite sampling) the equation:
\begin{equation}\label{eq:247}
x_g(\varepsilon)=\varepsilon-i\sum_{m\neq
0,m=-\infty}^{+\infty}\frac{(-1)^m}{4m\pi}\Phi(-4\pi m)\mathrm{e}^{i 4\pi
m\varepsilon}
\end{equation}
Deriving in $\varepsilon$, the Poisson identity gives:
\begin{equation}\label{eq:249}
\frac{\mathrm{d} x_g}{\mathrm{d}
\varepsilon}=\frac{1}{2}\sum_{l=-\infty}^{+\infty}\varphi(-\varepsilon+\frac{1}{4}-\frac{l}{2})
\end{equation}
The equation~\ref{eq:249} is a sum of infinite copies of
$\varphi(x)$ centered at 1/4 and 3/4 of each strip. It will be
used as a guide to select a response function p$(x)$ for the
simulations.

\subsection{Probability Distribution}

As we discussed in ref.~\cite{landi01}, given an identical signal
distribution for each event, the probability $\Gamma(x_g)$ to have
$x_g$ ( for each algorithm) is the product of the probability
P$(\varepsilon)$ to have $\varepsilon$ and $|\mathrm{d}\varepsilon(x_g)/\mathrm{d}
x_g|$:
\begin{equation}\label{eq:100}
  \Gamma(x_{g})=\mathrm{P}(\varepsilon)\Big|\frac{d\varepsilon(x_g)}{dx_{g}}\Big|
\end{equation}
This equation could be handled as a differential equation, if we
prove that the unknown function has always the same sign. We
showed in~\cite{landi01} that $d\varepsilon/dx_{g}$ is non
negative for a large set of signal distribution and algorithms
with different number of strips. For an infinite sampling,
equations~\ref{eq:242} and~\ref{eq:249} show immediately that $\mathrm{d}
x_g/\mathrm{d}\varepsilon$ is always non negative if
$\varphi(\varepsilon)$ is non negative.

In real events, the conditions giving equation~\ref{eq:100} are
always violated. The noise is present, the charge deposition
fluctuates, different strips are hit in different events, and we
never have a single $\varepsilon(x_g)$. The experimental
probability distribution is an average over different type of
signals produced by the particles and of responses of different
strips of the detector. Being the probability an average over
fluctuations, also the reconstruction will be an average, and it
will be effective if the fluctuations and the noise is not too
high.
With these limitations in mind and for a constant
P$(\varepsilon)$, the equation~(\ref{eq:100}) is easy to solve
given an initial condition:
\begin{equation}\label{eq:105}
\varepsilon(x_{g})=\tau\int_{x_g^0}^{x_{g}}\Gamma(y)dy+\varepsilon(x_g^0).
\end{equation}
The selection of the initial condition $\varepsilon(x_g^0)$ is a
critical point, we will see in the following, for $\theta\neq0$,
the amplitude of the systematic error given by the error on
$\varepsilon(x_g^0)$ .

To simplify, the symmetry of the strip response $p(x)$ will be
assumed. So, for symmetric $\varphi(x)$, two points have the
property $x_g(\varepsilon)=\varepsilon$: $x_g(\pm 1/2)=\pm 1/2$
and $x_g(0)=0$. In this case the initial condition is exactly
defined.
\subsection{The Data}
To fine tuning our simulation, we need a comparison with
experimental data. The data used are obtained from a test beam of
a tracker prototype for the PAMELA~\cite{PAMELA} experiment.

The beam particles are $\pi^+$ and the incidence angle is
orthogonal to the plane of the sensors. Each sensor is 300 $\mu$m
thick silicon wafer where the junction side and the ohmic side are
arranged to position measurements. In the junction side $p^+$
strips are implanted and one of two connected to the read out
electronics (pitch 50$\mu$m).  The ohmic side has $n^+$ strips
orthogonal to the strips of the junction side and each $n^+$
strips connected to the read out electronics (pitch 67 $\mu$m). No
magnetic field is present.

The sign of the read-out signals is arranged to be independent
from the collected charge. The events are selected in a standard
way and, after pedestal and common noise subtraction, they are
assembled with the data of the maximum-signal strip at the center
of an eleven component vector. The other vector components contain
the data of the neighboring strips. The pedestal and common noise
subtraction attenuates almost completely the ADC quantization.

Low signal values of any sign, contained in the data, are used in
the algorithms. No attention is given to the cluster size. We
select the events to have the sum of the collected signal in the
maximum strip and in its two adjacent greater than 40 ADC counts
and lower than 350 ADC counts. In this range of ADC counts the
Landau peak is fully contained. We cut the high energy tail where
it is probable to have $\delta$-rays or other reactions, and the
low energy tail for its low signal to noise ratio.

The far strip signals are use to estimate the noise and the
crosstalk.

The COG algorithms have the origin of their reference system in
the center of the strip with the maximum signal. (In section 6 we
will abandon this convention and the true position will be
considered)

We indicate with $x_{gn}$ the COG calculated with $n$ strips. The
two strip algorithm $x_{g2}$ has the best signal to noise ratio,
but, due to the crosstalk has an appreciable discontinuity in the
origin even with $\theta=0$ and regular strips. The noise strongly
attenuates the discontinuity, but  a drop around $x_{g2}\approx 0$
is evident in the probability distribution.

The $x_{g3}$ algorithm tends to have two region of low probability
at the strip borders.

The infinite sampling is excluded for the noise, thus we can
expect some deviation in the $\varphi(\varepsilon)$ reconstructed
with $\varepsilon$ extracted from  $x_{g2}$ and $x_{g3}$. These
deviations are well localized and we expect that $x_{g2}$ and
$x_{g3}$ are interesting reconstruction tools outside the regions
of strong deviations.

\subsection{Position Reconstruction}

In the $x_{g2}$ algorithm, we use maximum-signal strip and the
maximum (of any sign) between the two lateral. In the $x_{g3}$ we
use the maximum-signal strip and the two lateral.

 We assume that the probability P$(\varepsilon)$ is constant
on a strip length after the subtraction of the maximum integer
contained in $\varepsilon$, and the noise does not destroy the
validity of equation~\ref{eq:105}. The histograms of
$\Gamma(x_{g2})$ and $\Gamma(x_{g3})$ are inserted in
equation~\ref{eq:105}, and $\varepsilon_2(x_{g2})$ and
$\varepsilon_3(x_{g3})$ are obtained with a numerical integration.
The initial condition $\varepsilon_{2,3}(-1/2)=-1/2$ is used. Due
to the noise $\varepsilon_{2,3}(x_{g2,3})$  are an estimate of the
most probable $\varepsilon$ given $x_{g2,3}$. Where
$\varepsilon_{2,3}$ or $x_{g2,3}$ indicate simultaneously
$\varepsilon_{2}$ and $\varepsilon_{3}$ or $x_{g2}$ and $x_{g3}$.

The numerical forms of $\varepsilon_{2,3}(x_{g2,3})$ are not
practical for the successive uses, it is fundamental to have
analytical expressions. The natural way to construct analytical
expressions is to consider the periodicity of $\varepsilon(x_g)$
in presence of a true uniform illumination. This is not our case,
but we have to near this condition. Our set of data are scattered
on a large set of strips, and it is improbable to have the same
strip few times. In the algorithm, we collect together all the
signal on a single strip to generate a fictitious uniform
illumination on that strip, all the other strips have no
illumination. But, due to the uniformity of the strip properties,
we can consider all the strips with an identical incident
illumination and an identical probability $\Gamma(x_g)$, this
virtual total probability becomes periodical:
\[
\Gamma^p(x_g)=\sum_{j=-\infty}^{\infty}\Gamma(x_g+j)\, .
\]
With $\Gamma^p(x_g) $, the Fourier series (FS) is the natural way
to fit the function $\varepsilon_n-x_{gn}$ that is itself
periodic:
\begin{equation}
\begin{aligned}\label{eq:106}
  &\varepsilon_n(x_{gn})-x_{gn}=\sum_{k=-\infty}^{+\infty}\alpha_k \mathrm{e}^{i2\pi
  k\,  x_{gn}}\\
  &x_{gn}(\varepsilon_n)-\varepsilon_n=\sum_{k=-\infty}^{+\infty}\beta_k \mathrm{e}^{i 2\pi k\varepsilon_n}
\end{aligned}
\end{equation}
The periodicity of $\Gamma^p(x_{gn})$ is a method to handle the
events with $|x_{gn}|>1/2$.  In the $x_{g2}$-algorithm, it is very
improbable to have $|x_{g2}|>1/2$. The less energetic strip must
have a large negative value with its modulus greater than 2/3 the
energy of the other strip, and this never happens in our data. In
the $x_{g3}$ algorithm, events with $|x_{g3}|>1/2$ are possible,
and in the $x_{g4}$ algorithm an important fraction of events has
$|x_{g4}|>1/2$.

Given our interest in $\varepsilon_n(x_{gn})$, we use the first of
the equations~\ref{eq:106} with a reduced number of terms
($\alpha_k=0$ for $|k|\geq 15$) to filter the high frequencies of
the numerical integration on a small bin-size histogram. The
$\alpha_k$ are obtained with another numerical integration with
the standard form of the FS coefficients.

\begin{figure}[h!]
\begin{center}
\includegraphics[scale=0.7]{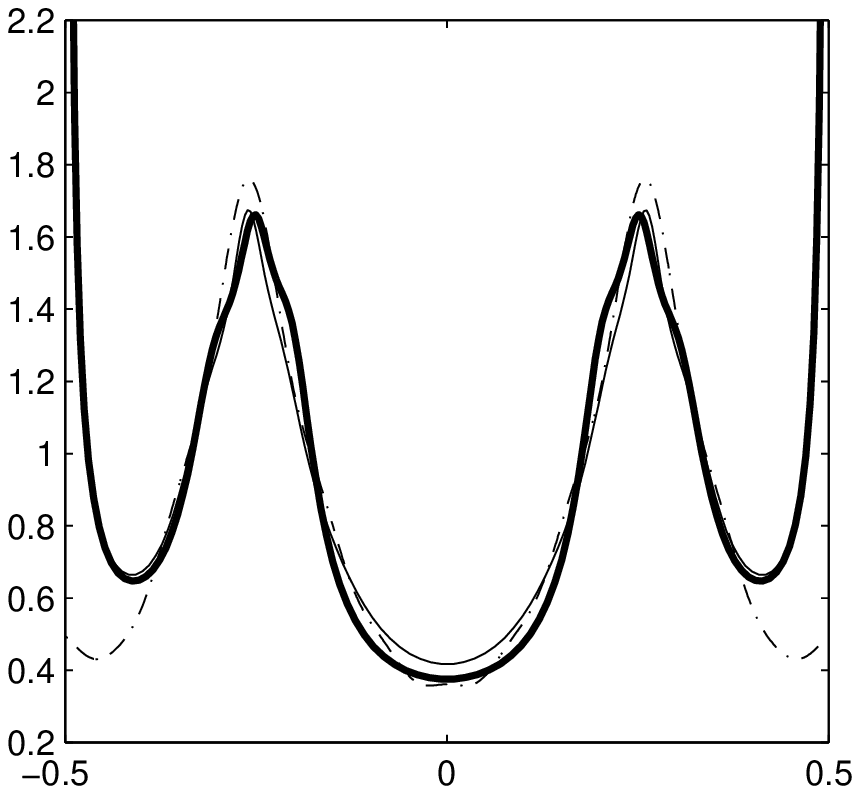}
\includegraphics[scale=0.7]{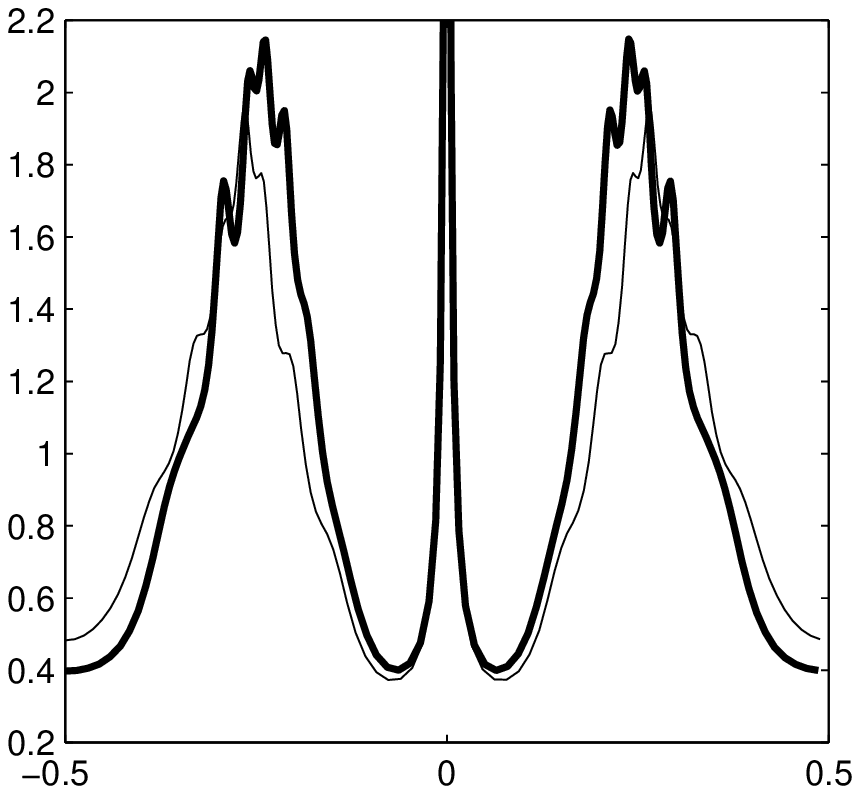}
\caption{\em To the left $\varphi(\varepsilon)$ obtained by the
$x_{g3}$ algorithm applied to the data (thick line), given by the
simulations (thin line), and deriving equation~\ref{eq:241}
(dash-dot line). To the right the algorithm is $x_{g2}$, the thick
line is given by the data, the thin line by the simulations.
}\label{fig:figura2}
\end{center}
\end{figure}

Dropping the number of strips in the notation, the derivative $\mathrm{d}
x_g(\varepsilon)/\mathrm{d} \varepsilon$  is given by the identity:
\begin{equation}\label{eq:107}
 \frac{\mathrm{d} x_g(\varepsilon)}{\mathrm{d} \varepsilon}=\frac{1}{\frac{\mathrm{d}
\varepsilon(x_g)}{\mathrm{d} x_g}}=[1+\sum_{k=-\infty}^{+\infty}(i 2\pi
k)\alpha_k \mathrm{e}^{i2\pi k x_g}]^{-1}.
\end{equation}
and the plot of the reconstructed $\varphi(\varepsilon)$ is
generated with $\varepsilon(x_g)$ on the $x$-axis, and the
$x_g$-function of equation~\ref{eq:107} on the $y$-axis. This
method is very effective,  as we can see in
figure~\ref{fig:figura2}. Here the $x_{g3}$ from data and from the
simulations are elaborated in the same way. From
equation~\ref{eq:249} we expect  two copies of
$\varphi(\varepsilon_3)$ centered at $\pm 0.25$. The similarity of
the two reconstructed $\varphi(\varepsilon_3)$ is noticeable and
they are near to the noiseless $\mathrm{d} x_{g3}(\varepsilon)/\mathrm{d}
\varepsilon$ (obtained deriving the equation~\ref{eq:241}). The
parameter $\alpha=0.0324$ is used in the simulations. The details
of the simulations will be given in the coming section, here we
test the response function and the diffusion parameter $\alpha$.

The two lateral increases are typical of the $x_{g3}$-algorithm.
In the noiseless case, two discontinuities are present in
$x_{g3}(\varepsilon_3)$ at $\varepsilon_3=\pm 1/2$. The noise
smooths the discontinuities in regions of fast variation that give
a substantial increases to the derivative. The noiseless
$x_{g3}(\varepsilon)$ and its derivative (given by the
equation~\ref{eq:241}) are defined in the interior of a single
strip and the two Dirac $\delta$ are outside the definition
domain.

To see the effect of limited sampling, in the left side of
figure~\ref{fig:figura2} the equation~\ref{eq:107} is applied to
$\varepsilon_2(x_{g2})$ given by the data and by the simulations.
Even here the two reconstructions are similar, and similar is the
number of oscillations, these are given by the lower number of
events respect to the regions with $\varepsilon\approx 0$ and
$\varepsilon\approx\pm 1/2$. The large increase for
$\varepsilon=0$ is typical of $x_{g2}$ that has there a
discontinuity in the absence of noise.

\begin{figure}[h!]
\begin{center}
\includegraphics[scale=0.7]{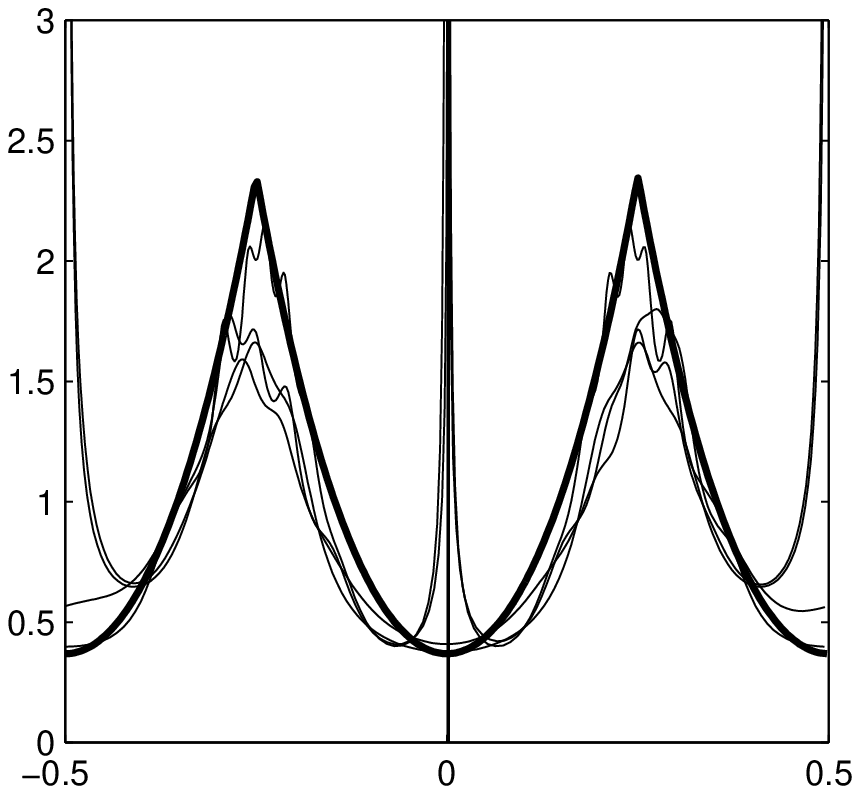}
\includegraphics[scale=0.7]{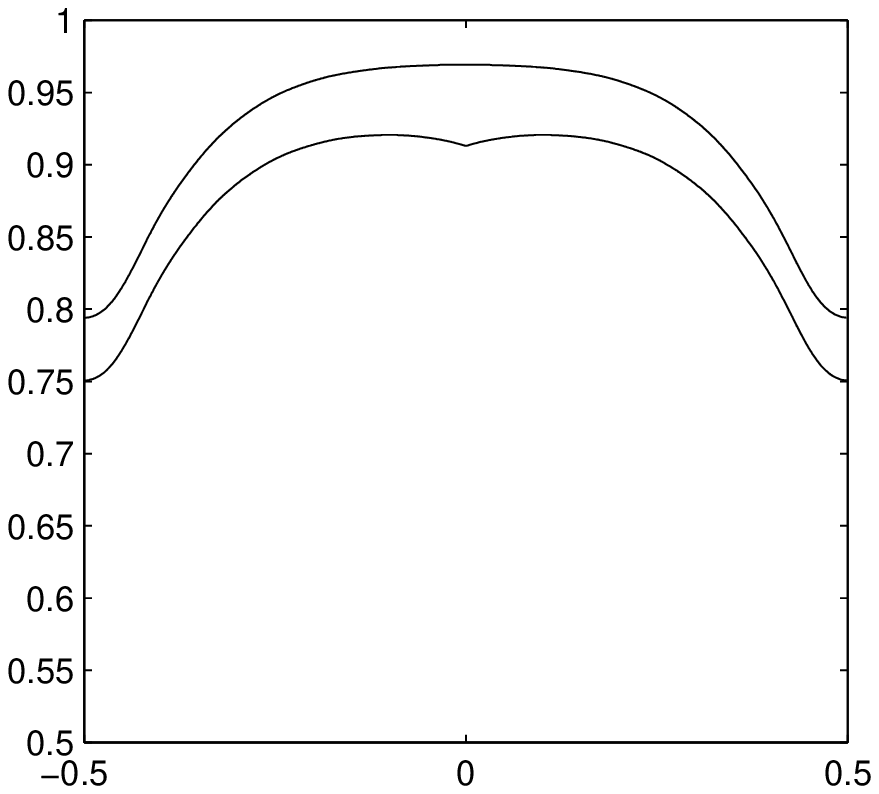}
\caption{\em The thick line is $\varphi(\varepsilon)$ as given by
equation~\ref{eq:249}, the thin lines are the other
reconstructions of figures~\ref{fig:figura2}. To the right, the
efficiency of $x_{g_3}$ (higher curve) and of $x_{g_2}$ (lower
curve). The signal loss is visible as the drop around $\pm 0.5$. }
\label{fig:figura53}
\end{center}
\end{figure}

The result of equation~\ref{eq:249} is reported in
figure~\ref{fig:figura53} with all the reconstructions of
figure~\ref{fig:figura2}.

\section{Simulations}

\subsection{Event Generation}


To generate the events, we use the following procedure. A set of
random points $\{\varepsilon_i\}$  is generated with an uniform
distribution on 3 strips. The points $\varepsilon_i$ are used to
calculate the fraction of energy $E_{\varepsilon_i}^l$
(equation~\ref{eq:241a}) for five strips $\{l\}$ centered around
the strip containing $\varepsilon_i$. For each event, a random
number is extracted with the distribution of the total
experimental charge (in ADC counts) collected by three strips (the
one with the maximum signal and the two adjacent). All the
$E_{\varepsilon_i}^l$ of the event $i$ are scaled with this random
factor. The noise is added to each strip with a gaussian
distribution modelled from the data.

\subsection{Losses}

Further details must be defined to fine tuning our simulations to
the data. From the beginning, we suppose that the signal shape is
the same at least on average. The noise and the fluctuations of
the primary ionization modify the pattern of the collected energy
for each events, but the $x_g$-histograms must not change. On the
contrary, the $x_g$-histograms for events with energy less than
the Landau maximum (175 ADC counts for our data) have a larger
probability for $|x_g|\approx 1/2$ than the events with higher
energies. An easy explanation of these differences can be a signal
loss in the detector  interior around the floating strip.

We can simulate this loss with an efficiency reduction around the
strip borders. Cutting two grooves around $x=\pm 1/2$ on $p(x)$,
one obtains the efficiency plots of figure~\ref{fig:figura53}. The
efficiency drop around $x=\pm 0.5$ is exclusively given by the
loss. This type of loss is qualitatively similar to that measured
in ref~\cite{krammer}; in our case it is guessed by the
$x_{g2,3}$-probability plots for events with different total
energy.

\subsection{Long Range Crosstalk}

\begin{figure}[h!]
\begin{center}
\includegraphics[scale=0.65]{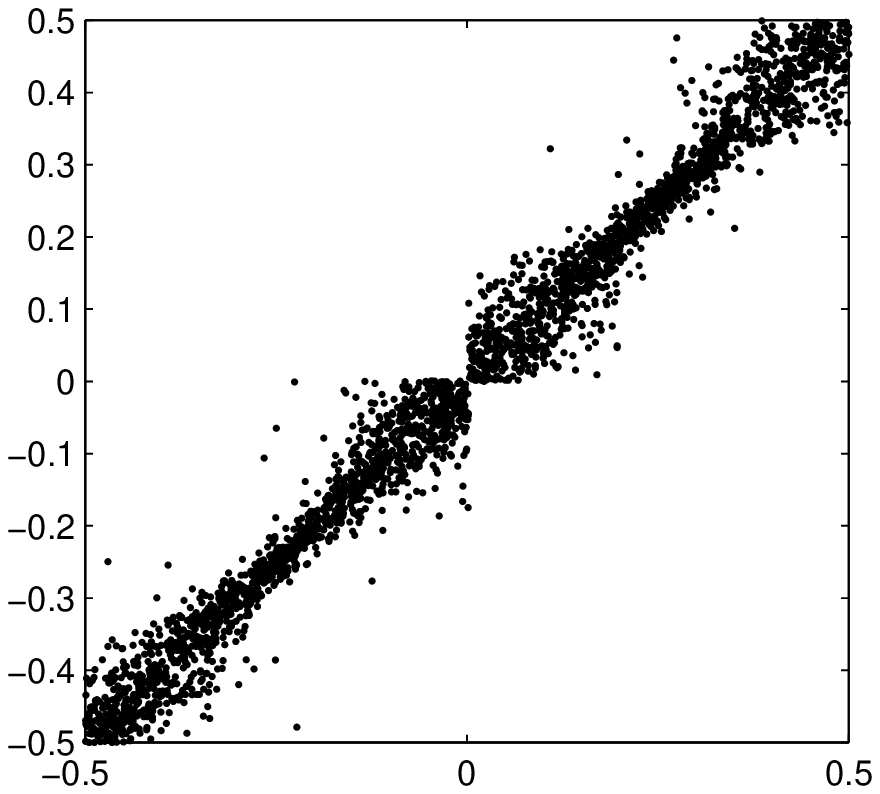}
\includegraphics[scale=0.65]{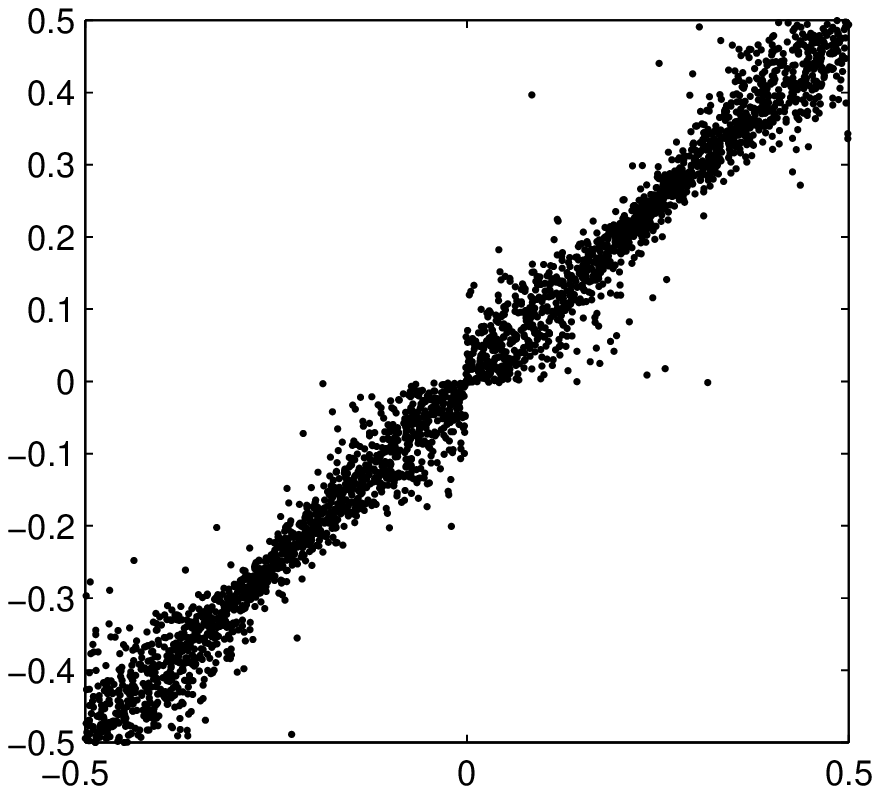}
\caption{\em Samples of
$\{\varepsilon_3(x_{g3}(j)),\varepsilon_2(x_{g2}(j))\}$. To the
left the data. To the right the simulations. Identical forbidden
regions are present in the two scatter plots.  }
\label{fig:figura22a}
\end{center}
\end{figure}

In section~\ref{sec:crosstalk} we discussed the main crosstalk
produced by the floating strip. But, in the strips adjacent to
that used in $x_{g3}$, we see an average shift of the collected
signals toward the maximum signal strip. This means that a long
range crosstalk is present in the data. To produce detailed
simulations, other terms must be added to $p(x)$ with tails longer
than that of figure~\ref{fig:figura1}. In
figure~\ref{fig:figura22a}, the effect of  the crosstalk (and the
quality of the simulations) is clearly seen. Here
$\varepsilon_3(x_{g3}(j))$ vs $\varepsilon_2(x_{g2}(j))$ for data
events and for simulated events can be compared, it is evident a
strict similarity of the distribution patterns. The total
elimination of all the crosstalk drastically modifies the scatter
plot as can be see in the plots of figure~\ref{fig:figura64}.
\subsection{Data and Simulations}\label{sec:datasim}

\begin{figure}[h!]
\begin{center}
\includegraphics[scale=0.7]{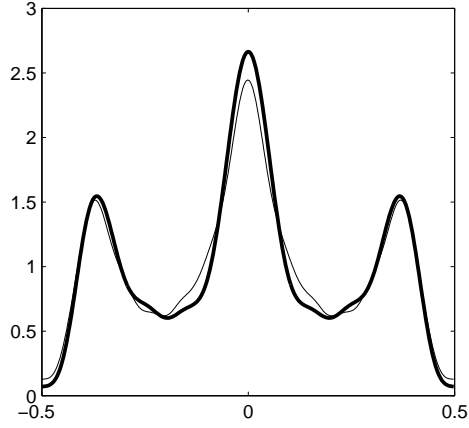}
\caption{\em Probability distribution $\Gamma^p(x_{g3})$ of the
simulation (thin line) and of the data (thick line).
 }\label{fig:figura2b}
\end{center}
\end{figure}

The consistency of the simulations with the data can be
appreciated in figure~\ref{fig:figura2b}. Here the probability
distributions of $\Gamma^p x_{g3}$ are compared. The two curves
are obtained from equation~\ref{eq:106}, but this time we plot
$\mathrm{d}\varepsilon(x_{g3})/\mathrm{d} x_{g3}$ versus $x_{g3}$. This is
apparently a trivial operation, first $\Gamma^p(x_{g3})$ is
integrated an after it is derived to return to $\Gamma^p(x_{g3})$.
In reality this procedure applies a consistent low pass filter to
$\Gamma^p(x_{g3})$.

\begin{figure}[h!]
\begin{center}
\includegraphics[scale=0.65]{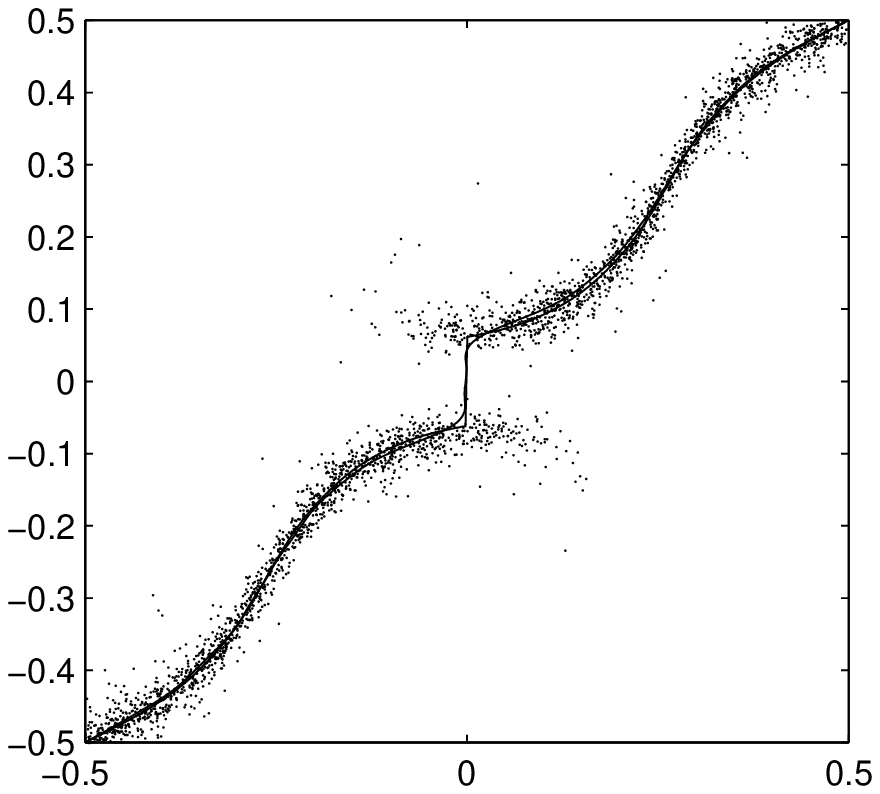}
\includegraphics[scale=0.65]{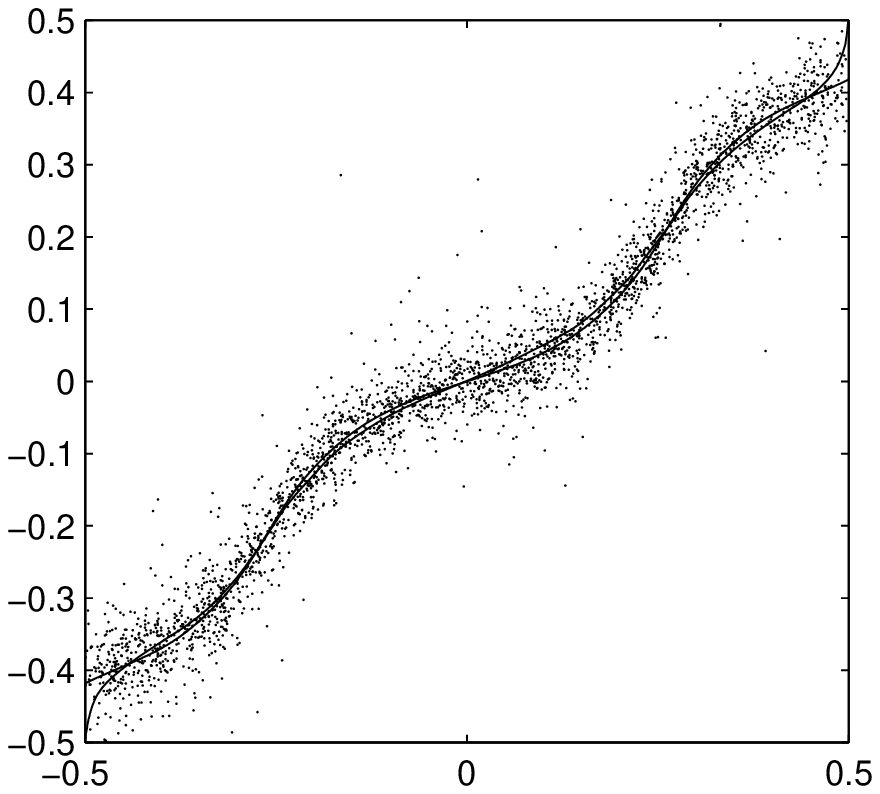}
\caption{\em To the left, the $x_{g2}$-positions in function of
the corresponding $\varepsilon$ for the simulated events. The two
overlapping curves are: $x_{g2}(\varepsilon_2)$ and the noiseless
$x_{g2}(\varepsilon)$. To the right, $x_{g3}$ and $\varepsilon$
for the simulated events. The two overlapping curves are $
x_{g3}(\varepsilon_3)$ and the noiseless $ x_{g3}(\varepsilon)$.
 } \label{fig:figura21}
\end{center}
\end{figure}

Figure~\ref{fig:figura21} reports a sample of simulated $x_{g2}$
and $x_{g3}$ values. The two overlapping lines are
$x_{g2}(\varepsilon_2)$ obtained from simulated events with noise
and the noiseless $x_{g2}(\varepsilon)$. If we exclude a smoothing
in the rapid variation regions (around $x_{g2}=0$), the overlap is
almost complete. The noise works as a low pass filter. Similar
overlap is observed even for the noisy $x_{g3}(\varepsilon_3)$)
with the plot of noiseless $x_{g3}(\varepsilon)$). In spite of
these overlaps, the error
$\varepsilon_{2,3}(x_{g2,3})-\varepsilon$ remains appreciable. In
fact, the reconstruction algorithm $\varepsilon_{2,3}(x_{g2,3})$
explores the scatter-plot of figure~\ref{fig:figura21} along the
horizontal line going through $x_{g2,3}$. The noise shifts
$x_{g2,3}$ along a vertical line passing through the noiseless
$x_{g2,3}(\varepsilon)$. If the shift is large and/or the slope of
$x_{g2,3}(\varepsilon_{2,3})$ is small the error is large.

The easy agreement of the simulations with the data (and the small
error of the $\varphi(x)$-reconstructions) allows to state that
the key point of the simulations for the detector with floating
strips is a response function $p(x)$ with the form of
figure~\ref{fig:figura1}. This $p(x)$ gives even an explanation of
the better quality of the reconstruction (and of its error almost
independent from $\theta$, as it will be seen in the following).
In fact, for signal distributions of the size used, its effect is
similar to a triangular $p(x)$ with base $2\tau$. As proved in
ref.~\cite{landi01}, the triangular function with base $2\tau$ is
one of the response function that are free from the COG
discretizzation error for any signal distribution (ideal
detector). The noise and the information loss, due to a restricted
number of strips used in the algorithm, are the most significant
limiting factors in the reconstruction, and a true triangular
$p(x)$ gives  modest improvements.

\section{Normal-strip side}

\subsection{The Response Function}

The ohmic-side of the sensor has no floating strips, we will call
normal strip side the ohmic-side, and normal strips its strips.

The lack of a capacitive coupling in this side produces a large
reconstruction error. For $\theta=0^\circ$, an important fraction
of the signal tends to concentrate on a single strip giving an
high probability to have $x_{g2,3}\approx 0$. A small interstrip
capacity is present, but its crosstalk effect remains modest. At
any rate, to produce a good simulation, we have to account for
this crosstalk with the following empirical response function
$p(x)$:
\begin{equation}\label{eq:260}
\begin{aligned}
  p(x)=&\int_{-\infty}^{+\infty}\mathrm{d} x'\,\Pi(x-x')\Big\{0.91\delta
  (x')+0.063[\delta(x'-\tau/2)+\delta(x'+\tau/2)]\\
  &+0.027[\delta(x'-3\tau/2)+\delta(x'+ 3 \tau / 2)]\Big\}
\end{aligned}
\end{equation}
The main part of $p(x)$ is given by $0.91\,\Pi(x)$ (i.e., an
interval function), and the small terms are selected to produce an
uniform crosstalk. $p(x)$ is fine tuned to produce a good
agreement of the histograms (and cross-correlations) obtained from
the simulation with the corresponding obtained from the data.

\begin{figure}[h!]
\begin{center}
\includegraphics[scale=0.8]{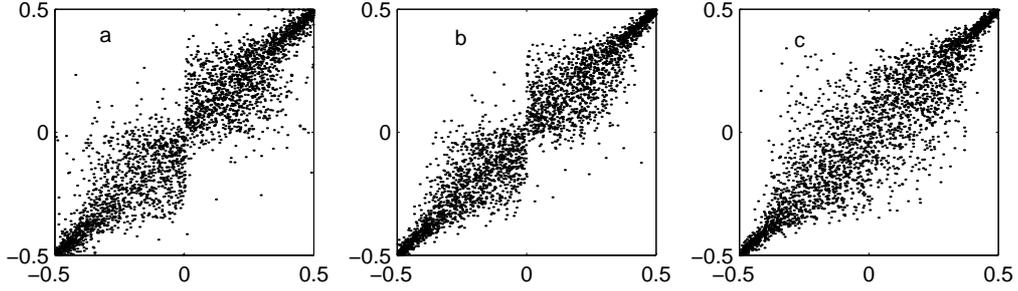}
\caption{\em Samples of
$\{\varepsilon_3(x_{g3}(j)),\varepsilon_2(x_{g2}(j))\}$. Figure a)
for the data, b) the simulations with crosstalk, and c) the
simulations without the crosstalk .
 } \label{fig:figura64}
\end{center}
\end{figure}

In figure~\ref{fig:figura64}, we show the correlation of
$\varepsilon_3(x_{g3}(j))$ vs. $\varepsilon_2(x_{g2}(j))$ in the
data, in the simulation with the crosstalk of
equation~\ref{eq:260}, and without crosstalk ($p(x)=\Pi(x)$). As
it is clearly seen, the absence of crosstalk produces a plot
drastically different from the data. The crosstalk of
equation~\ref{eq:260} gives correlations with exclusion regions
very similar to the experimental one. This test looks more
sensitive than the histograms, here even the noise model can be
compared.

\subsection{Simulations}

The simulated events are generated as in the case of the floating
strip, with some evident modifications. The parameter $\alpha$ is
taken to be $0.0289$ (partially scaled to keep into account the
difference of $\tau=67\mu$m), and $p(x)$ is given by
equation~\ref{eq:260}.

\begin{figure}[h!]
\begin{center}
\includegraphics[scale=0.7]{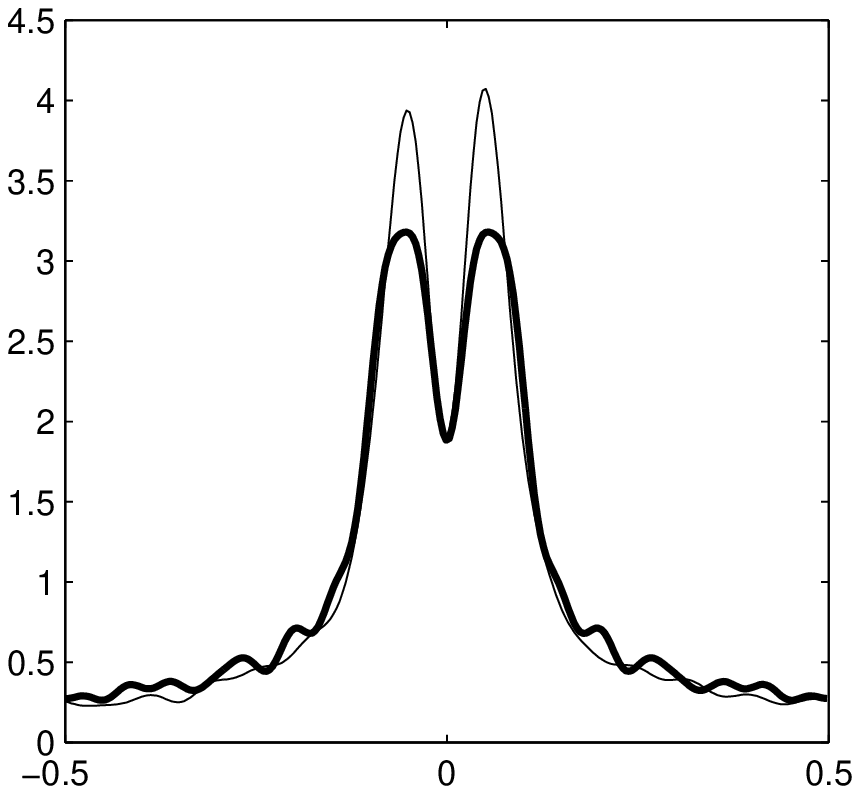}
\includegraphics[scale=0.7]{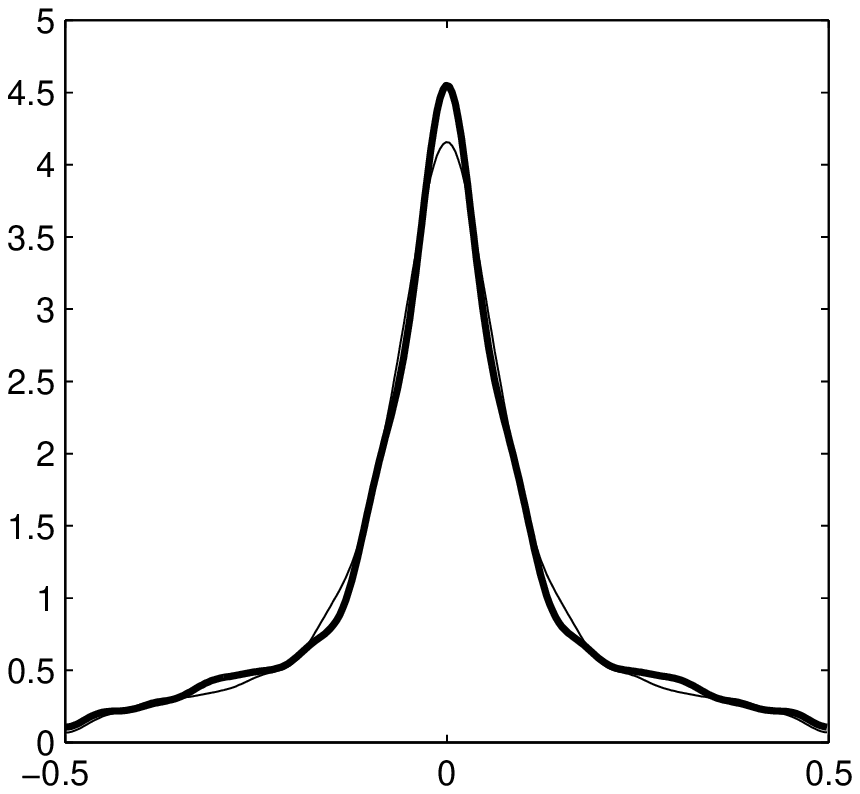}
\caption{\em To the left, probability distribution for $x_{g2}$
from the simulations (thin line), and from the data (thick line).
To the right, the probability distribution for $x_{g3}$;
simulations (thin line), data (thick line). } \label{fig:figura26}
\end{center}
\end{figure}

\begin{figure}[h!]
\begin{center}
\includegraphics[scale=0.7]{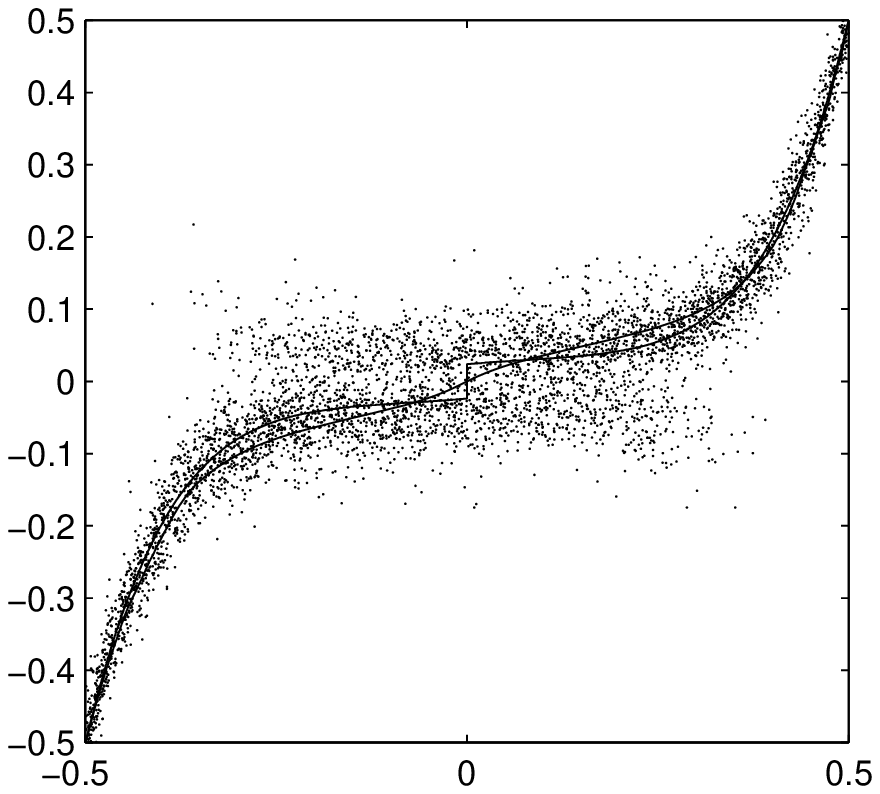}
\includegraphics[scale=0.7]{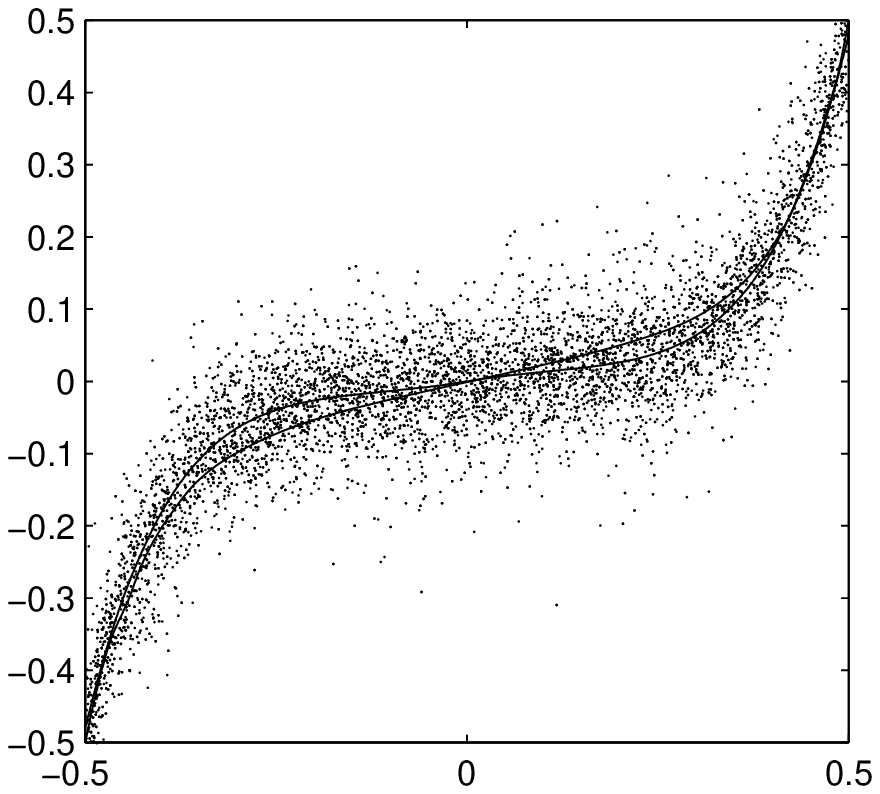}
\caption{\em To the left, a sample of
$\{\varepsilon(j),x_{g2}(j)\}$, the two curves are: the
discontinuous one is the noiseless $x_{g2}(\varepsilon)$, the
other is $x_{g2}(\varepsilon_2)$ with $\varepsilon_2$ given by
equation~\ref{eq:105}. To the right, a sample of
$\{\varepsilon(j),x_{g3}(j)\}$. The two curves are the noiseless
$x_{g3}(\varepsilon)$ and $x_{g3}(\varepsilon_3)$}
\label{fig:figura33}
\end{center}
\end{figure}

A set of random $\varepsilon_i$ values is generated with an
uniform distribution over three strips. The $\varepsilon_i$-values
are the COG of initial ionization distributions. The fraction of
the charge collected by each strip is calculated with
equation~\ref{eq:241a}, and the simulated collected charge is
obtained multiplying by a random factor with a distribution
similar to that in the data. A fluctuation with a Poisson
distribution  is added to the charge collected in each strip. The
additive noise is gaussian with zero mean and standard deviation
similar to the noise distribution of the distant strips in the
data. A comparison of the probability distributions for $x_{g2}$
and $x_{g3}$ is reported in figure~\ref{fig:figura26}. The plots
are produced as explained in section~\ref{sec:datasim} for the
floating strip side. Even in this case the agreement is excellent.
Figure~\ref{fig:figura33} is the equivalent of
figure~\ref{fig:figura21} for the normal strip side. Here the
simulated events are scattered far from the curves
$x_{g2,3}(\varepsilon_{2,3})$ or the noiseless
$x_{g2,3}(\varepsilon)$, and the position reconstruction with
$\varepsilon_{2,3}(x_{g2,3})$ has a larger error compared to the
floating strip side.  The overlaps of
$x_{g2,3}(\varepsilon_{2,3})$ with the noiseless
$x_{g2,3}(\varepsilon)$ are smaller than these for the floating
strips.

\section{Directions with $\theta\neq 0$}
\subsection{Simulations}

The angle $\theta$ is defined in figure~\ref{fig:figura0}, and
$L_x$ is given by $L_x=L_0\tan(\theta)$. With $L_x\neq 0$, the
different diffusion times of the top and the bottom of the track
are shifted, and $\varphi(x)$ becomes asymmetric.
Figure~\ref{fig:figura1*36} shows these asymmetries.

For $\theta\neq 0$ we have no data consistent with these used for
$\theta=0$, and we cannot test the simulations. But, given the
agreement reached for $\theta=0$, it is reasonable to proceed with
our simulations at least in the range $0^\circ\leq\theta\leq
20^\circ$. At any rate, we will prove with analytical development
our critical results delineated by the simulations. The
$\theta$-range is covered with steps of $1^\circ$.

For $\theta>20^\circ$, the loss of information of $x_{g2}$ and
$x_{g3}$ starts to produce non optimal results. Algorithms with a
greater number of strips must be explored and the simulations
should be calibrated with the data. Our use of $x_{g4}$ will be
limited to correct the integration constant of
equation~\ref{eq:105}.

In this section $\varepsilon_n$ and $x_{gn}$ indicate the full
reconstructed positions in the absolute reference system of the
sensor. The origin of this reference system is in the center of a
strip and the positions of the maximum signal strip is always an
integer number.

\subsection{Standard Deviations for the Reconstructions (Normal Strips)}

\begin{figure}[h!]
\begin{center}
\includegraphics[scale=0.8]{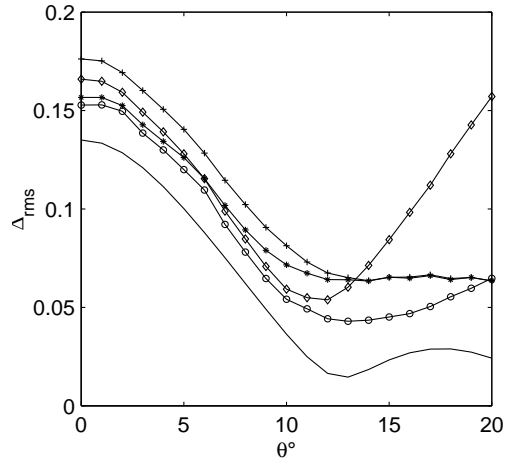}
\caption{\em Standard deviations for various reconstructions,
diamond $x_{g2}$, circle $\varepsilon_2$, crosses $x_{g3}$,
asterisks $\varepsilon_3$, continuous line the results of
equation~\ref{eq:262}}\label{fig:figura37}
\end{center}
\end{figure}

The simulations allow to calculate the standard deviation (SD) in
function of $\theta$ of the various reconstructions, we will start
with the normal strips. The trends of the SD can be compared with
the COG error for an infinite sampling of noiseless signal
distribution. The mean square error for uniform response function
was calculated in ref.~\cite{landi02} for two dimensional systems.
The reduction to one dimensional systems is trivial and gives:
\begin{equation}\label{eq:262}
\int_{-1/2}^{+1/2}(x_g(\varepsilon)-\varepsilon)^2\mathrm{d}\varepsilon=
\sum_{k=1}^{+\infty}|\Phi(-2k\pi)|^2(P'(-2k\pi))^2
\end{equation}
Figure~\ref{fig:figura37} reports various plots of SDs. The
continuous line is the result of equation~\ref{eq:262} at various
$\theta$. The other curves are the SD of
$\varepsilon_{2,3}(x_{g2,3})-\varepsilon$ and of
$x_{g2,3}-\varepsilon$. For small $\theta$ the line of
 is not far from  that of
$\varepsilon_{2}(x_{g2})-\varepsilon$. For $\theta>12^\circ$ the
SD of $x_{g2}-\varepsilon$ starts to increase abandoning the
common trends of the other reconstructions. The origin of this
increase is connected to the drastic loss of information of the
two strip algorithm that produces a large set of forbidden values
around $x_{g2}\approx 0$. In spite of this, the $x_{g2}$-algorithm
gives an excellent reconstruction $\varepsilon_{2}(x_{g2})$ with a
SD better than the SD of $\varepsilon_{3}(x_{g3})-\varepsilon$.

A sort of saturation effect appears when $\theta>12^\circ$ for
$\varepsilon_{2}(x_{g2})-\varepsilon$,
$\varepsilon_{3}(x_{g3})-\varepsilon$ and $x_{g3}-\varepsilon$.
Part of the trend of the analytical calculation~\ref{eq:262} is
present in the reconstruction errors up to $\theta\approx
12^\circ$, beyond this value the noise (and the loss of
information) dominates and the algorithms tend to loose their
relations to the noiseless lossless model.

\subsection{ Distributions of $\varepsilon_{2,3}-\varepsilon$}

\begin{figure}[h!]
\begin{center}
\includegraphics[scale=0.65]{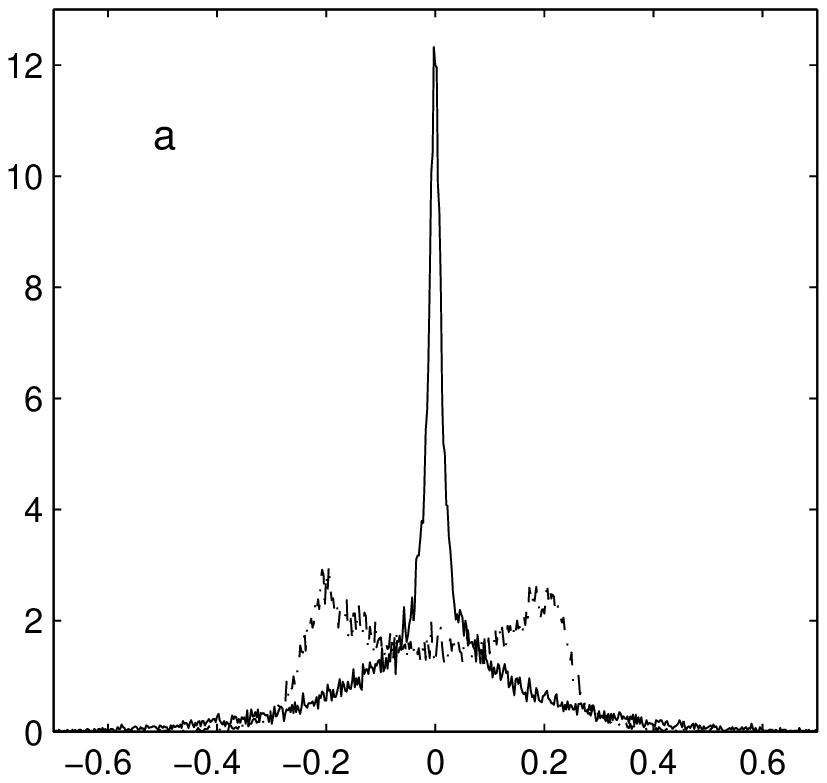}
\includegraphics[scale=0.65]{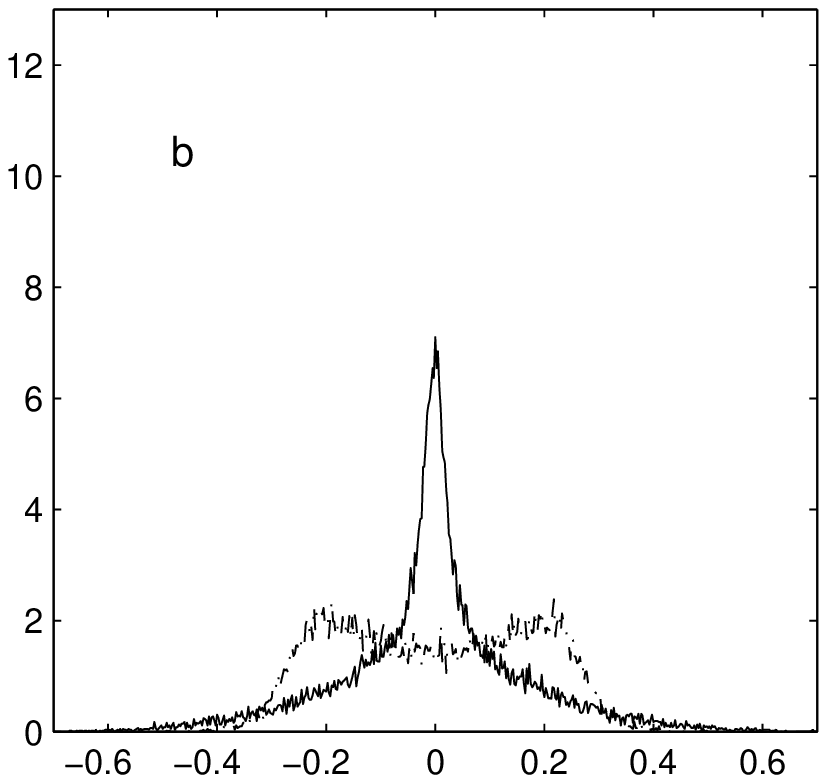}
\includegraphics[scale=0.65]{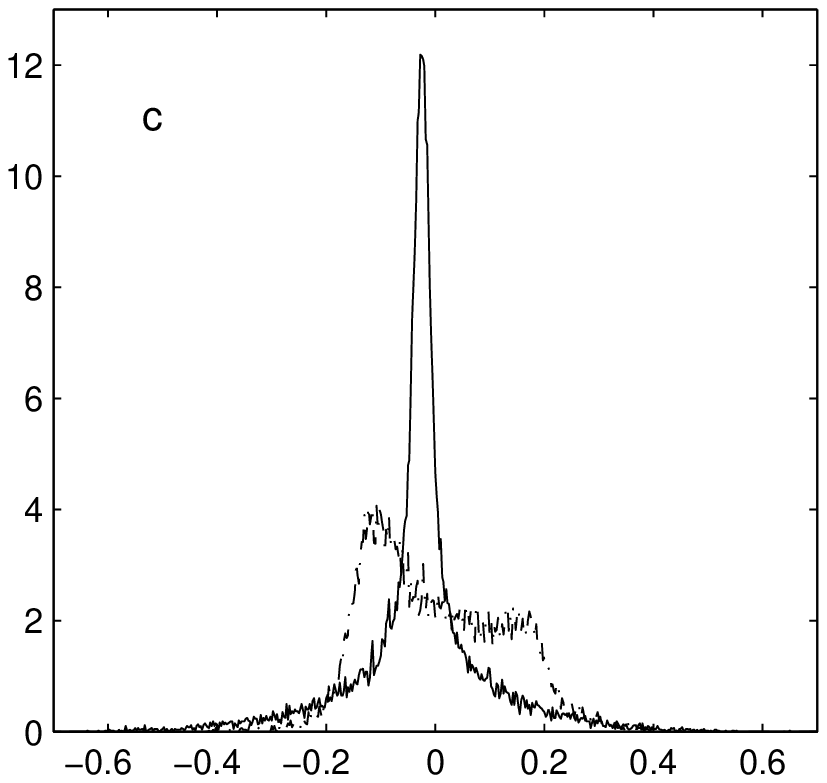}
\includegraphics[scale=0.65]{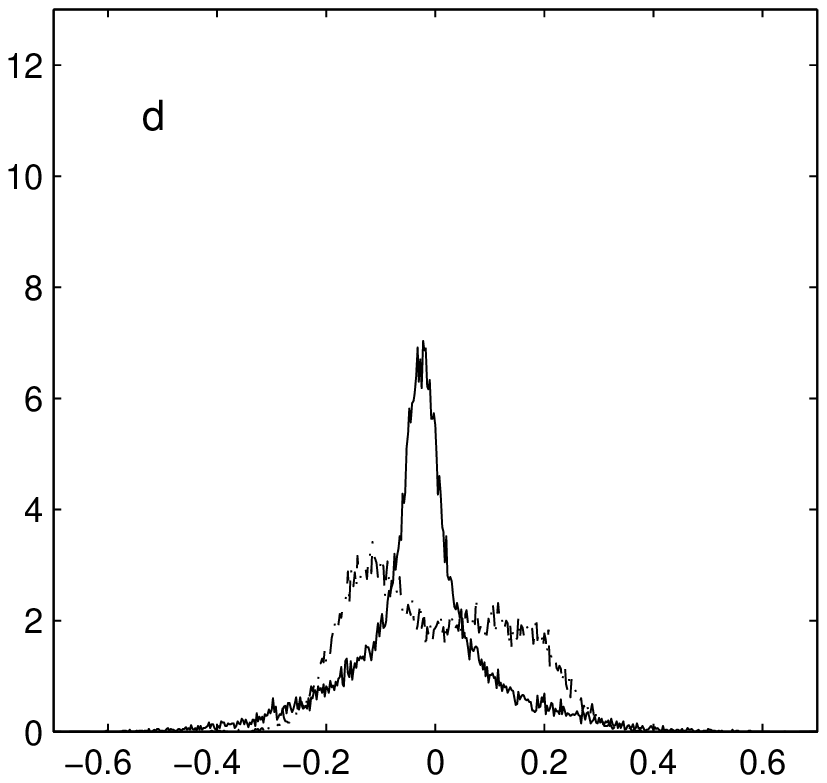}
\caption{\em Figure (a) is the histogram of
$\varepsilon_2(x_{g2})-\varepsilon$ (continuous line) and of
$x_{g2}-\varepsilon$ (dash-dotted line) for normal strips and
$\theta=0$. (b) is the histogram of
$\varepsilon_3(x_{g3})-\varepsilon$ (continuous line) and of
$x_{g3}-\varepsilon$ (dash-dotted line). (c) and (d) are for
$\theta=5^\circ$. In (c) it is evident the non zero position of
the maximum. } \label{fig:figura68}
\end{center}
\end{figure}

The SD's are not good parameters to characterize the differences
$\varepsilon_{2,3}(x_{g2,3})-\varepsilon$. Interesting plots are
the histograms of the probabilities of
$\varepsilon_{2,3}(x_{g2,3})-\varepsilon$ and of
$x_{g2,3}-\varepsilon$. We report them for $\theta=0^\circ$ and
$\theta=5^\circ$ in the figures~\ref{fig:figura68}. The
differences $\varepsilon_{2,3}(x_{g2,3})-\varepsilon$ have
distributions similar to the Cauchy probability density with a
small full width at half maximum (FWHM). It is interesting the
dramatic reduction of the FWHM of
$\varepsilon_{2,3}(x_{g2,3})-\varepsilon$ respect to the FWHM of
$x_{g2,3}-\varepsilon$ al low $\theta$. No comparable reduction is
observed in the SDs that, excluding the SD of
$x_{g2}-\varepsilon$, have values and trends similar. The FWHM is
an important parameter to describe these distributions. At
increasing $\theta$, the distributions slowly tend to gaussians.

\subsection{Error Due to $\varphi(x)$ Asymmetry}
In figure~\ref{fig:figura68} we start to see the effect of the
asymmetry of $\varphi(x)$ in the integration constant of
equation~\ref{eq:105}. For $\theta=5^\circ$, the maximum of the
distributions for $\varepsilon_{2,3}(x_{g2,3})-\varepsilon$ is
shifted from zero, and all the reconstructed values are shifted
accordingly. Even if modest in the scale of the SDs, this shift is
of the same order of the FWHM. For the floating strip sensor, the
shift can be of the order of the SD, as we will see in the
following.
\begin{figure}[h!]
\begin{center}
\includegraphics[scale=0.7]{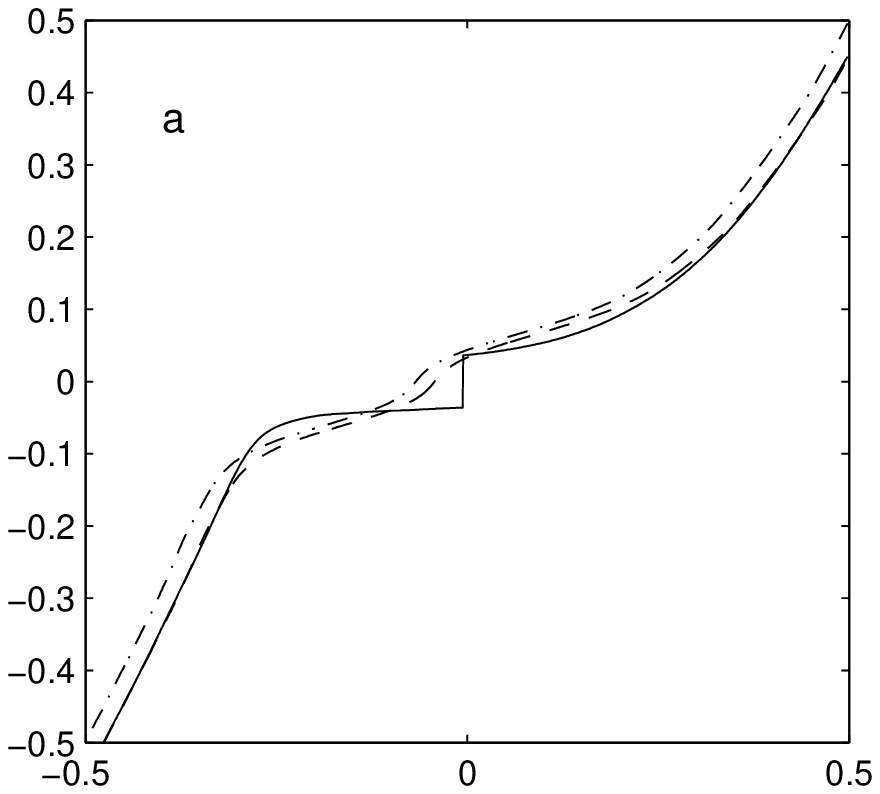}
\includegraphics[scale=0.7]{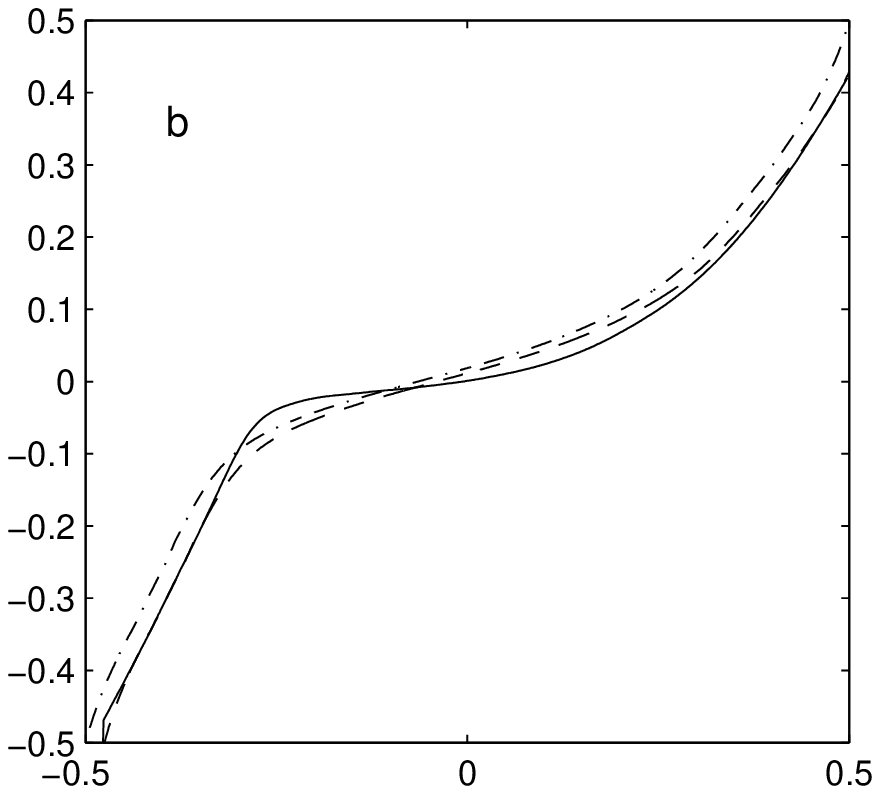}
\caption{\em To the left (a) is the plot of $x_{g2}$ vs
$\varepsilon_2(x_{g2})$ (dash-dot line) and $x_{g2}$ noiseless vs.
$\varepsilon$ (solid line) for $\theta=5^o$, the two curves do not
coincide as in figure~\ref{fig:figura21}, the dashed line is
$x_{g2}$ vs $\varepsilon_2^r(x_{g2})$ corrected as explained in
the following. This curve has large overlaps with the noiseless
one. To the right (b) the plots of $x_{g3}$ versus
$\varepsilon_3(x_{g3})$ (dash dot line) and of $x_{g3}$ noiseless
versus $\varepsilon$ (solid line) and $x_{g3}$ vs
$\varepsilon_3^r(x_{g3})$ (dashed line)} \label{fig:figura70}
\end{center}
\end{figure}

Figure~\ref{fig:figura70} illustrates the origin of the shift in
the maximum of the $\varepsilon_{2,3}(x_{g2,3})-\varepsilon$
probabilities. The lines $x_{g2,3}(\varepsilon_{2,3})$ do not
overlap with the noiseless $x_{g2,3}(\varepsilon)$, as we have for
$\theta=0$, they are shifted to right by a small quantity, thus
the reconstructed values $\varepsilon_{2,3}(x_{g2,3})$ of the
events, the maximum and the mean values of
$\varepsilon_{2,3}(x_{g2,3})-\varepsilon$. The corrections, we are
going to define, will rigenerate the overlaps. Having to work with
mean values, definitions are necessary:
\begin{equation}
\begin{aligned}
&M\varepsilon_{2,3}=\sum_{j=1}^N\frac{\varepsilon_{2,3}(x_{g2,3}(j))-\varepsilon(j)}{N}\\
&M_{x2,3,4}^l=\sum_{j=1}^N\frac{x_{g2,3,4}(j)-\mu(j)}{N},
\end{aligned}
\end{equation}
where $\mu(j)$ is the position of the center of the strip with the
maximum signal for the event $j$. The term $x_{gn}(j)-\mu(j)$ is
the COG of the event $j$ in the virtual strip we used to calculate
the probability distribution. We call local these averages to
stress their independence from the true event position.
\begin{figure}[h!]
\begin{center}
\includegraphics[scale=0.7]{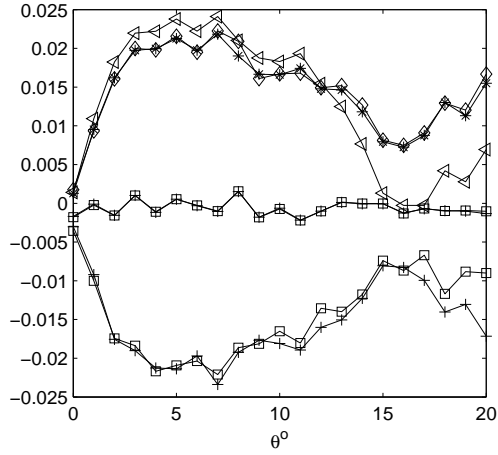}
\caption{\em Plots of mean values in function of $\theta$.
Diamonds $M_{x4}^l$, asterisks $M_{x3}^l$, triangles $M_{x2}^l$,
crosses $M\varepsilon_2$, squares $M\varepsilon_3$, the lines
around zero are $M\varepsilon_3^r$ and $M\varepsilon_2^r$
}\label{fig:figura60}
\end{center}
\end{figure}

Figure~\ref{fig:figura60} summarizes the $\theta$-dependent
averages: $M\varepsilon_2$, $M\varepsilon_3$, $M_{x 2}^l$,
$M_{x3}^l$ and $M_{x 4}^l$. The figure is complex to read in
detail. But, at first glance, two almost symmetric group of lines
respect to zero are evident. The lines for negative values are the
systematic errors $M\varepsilon_2$ and $M\varepsilon_3$. The lines
for positive values are the averages $M_{x 2}^l$, $M_{x3}^l$ and
$M_{x 4}^l$. The lines (overlapped) around zero are the corrected
$M\varepsilon_2$ and $M\varepsilon_3$. It easy to guess that the
sought corrections are near to $M_{x 2}^l$, $M_{x3}^l$ or $M_{x
4}^l$, in fact, the sum of $M\varepsilon_2$ or $M\varepsilon_3$
with $M_{x 3}^l$ or $M_{x 4}^l$ is around zero. A careful
observation shows that $M_{x 2}^l$ is not a good correction.

Given these empirical evidences, it is important to demonstrate
their origin to assure us that no simulation inconsistencies mimic
the rule. With this aim, we must go in depth in the reconstruction
algorithms to isolate the constants (in average) they contain.
\subsection{The Complete $\varphi(x)$-Reconstruction Theorem}

First of all, we must demonstrate the last element required to
complete the reconstruction theorem that evidently emerges from
equations~\ref{eq:242} and \ref{eq:105}.
The steps of this reconstruction are recalled here.

Given a set of noiseless and lossless events, originating by the same
$\varphi(\varepsilon)$, the COG $x_g$ (infinite sampling) for
each event is calculated and a probability distribution
$\Gamma(x_g)$ is obtained.

The events are supposed to uniformly illuminate all the strips,
and a periodic $\Gamma^p(x_g)$ ($\Gamma^p(x_g)=\Gamma^p(x_g+L) \
\forall L \in \mathbb{Z} $) is generated.

The integration of $\Gamma^p(x_g)$ gives:

\begin{equation}\label{eq:265a}
\varepsilon(x_{g})=\int_{x_g^0}^{x_{g}}\Gamma^p(y)dy+\varepsilon(x_g^0).
\end{equation}
Equation~\ref{eq:265a} produces the functions $\varepsilon(x_g)$
or $x_g(\varepsilon) $. The periodicity of $\Gamma^p(x_g)$ allows
the form~\ref{eq:106} of a FS for their differences with the
following expressions for $\alpha_n$ and $\beta_k$:
\begin{equation}
\begin{aligned}\label{eq:265b}
  &\alpha_n=\int_{-1/2}^{+1/2}\mathrm{e}^{-i2\pi n y}(\varepsilon(y)-y)\mathrm{d} y\\
  &\beta_k=\int_{-1/2}^{1/2}(y-\varepsilon(y))\mathrm{e}^{-i2\pi k\varepsilon(y)}\Gamma(y)\mathrm{d} y
\end{aligned}
\end{equation}

For a normalized non negative $\varphi(\varepsilon)$ (
$\varphi(\varepsilon)=\Pi(\varepsilon)\,\varphi(\varepsilon)$) and
the response function $p(x)=\Pi(x)$, the derivate $\mathrm{d}
x_g(\varepsilon)/\mathrm{d}\varepsilon$ allows the selection of the
function $\varphi(\varepsilon-1/2)$ :
\begin{equation}\label{eq:265c}
\Pi(\varepsilon-1/2)\frac{dx_{g}}{d\varepsilon}=
\varphi(\varepsilon-1/2).
\end{equation}
The limitation of the function $\varphi(\varepsilon)$ to have the
support less than a strip can be easily overcome given
$p(x)=\Pi(x)$. Assembling various strips the support can be
extended. This method cannot be used with the response function
used in equation~\ref{eq:249}. Even in this case the function
$\varphi(\varepsilon)$ can be reconstructed, but its support must
be less than half strip.

The equations~\ref{eq:265a} and~\ref{eq:265c} contain the
integration constant $\varepsilon(-1/2)$. In the simulations,
$\varepsilon(-1/2)$  is always set equal to $-1/2$, this position
is exact for symmetric $\varphi(\varepsilon)$. For asymmetric
$\varphi(\varepsilon)$ the incorrect $\varepsilon(-1/2)$ shifts
$\varphi(\varepsilon)$ by $\varepsilon(-1/2)+1/2$ giving an
incorrect reconstruction. The exact value $\varepsilon(-1/2)$ must
be extracted from the data to extend the reconstruction theorem to
asymmetrical $\varphi(\varepsilon)$.

Identical reasoning can be applied to the conventional $\eta$
algorithm, here the integration starts from zero, the position
$\varepsilon(0)=0$ is correct for $\theta=0$ and symmetric
$\varphi(\varepsilon)$. For $\theta\neq 0$ a
non zero constant is required.

The empirical property, evident in figure~\ref{fig:figura60},
gives a track to complete the theorem. Due to our definitions, we
can write the following relations for the mean value of $x_g$:
\begin{equation}\label{eq:290}
\int_{-1/2}^{+1/2}\Gamma^p(x_g) x_g \mathrm{d} x_g=\int_{-1/2}^{+1/2}x_g
\frac{\mathrm{d}\varepsilon(x_g)}{\mathrm{d} x_g}  \mathrm{d} x_g=
\int_{\varepsilon(-1/2)}^{\varepsilon(+1/2)} x_g(\varepsilon) \mathrm{d}
\varepsilon
\end{equation}%
Our assumption about the reconstruction of noiseless and lossless
events gives for  $x_g(\varepsilon)$ the equation:
\begin{equation}\label{eq:293}
x_{g}(\varepsilon)=\varepsilon+i \sum_{k\neq 0,
k=-\infty}^{+\infty}\Phi(-2k\pi)\mathrm{P}'(-2k\pi) \exp(i\,
2k\pi\varepsilon)
\end{equation}
for a generic uniform and symmetric response function $p(x)$, and
any normalized $\varphi(\varepsilon)$. For non symmetric $p(x)$ a
constant term must be added.

A fundamental property of equation~\ref{eq:293} is the absence of
the $k=0$-term in the sum of exponentials (FS). The integral of
$x_g(\varepsilon)$ on a strip length (the period of the FS) has no
contribution from the FS, and equation~\ref{eq:290} becomes:
\begin{equation}
\int_{\varepsilon(-1/2)}^{\varepsilon(+1/2)} x_g(\varepsilon) \mathrm{d}
\varepsilon=\frac{1}{2}[\varepsilon(1/2)^2-\varepsilon(-1/2)^2]
\end{equation}
From the periodicity of $x_g(\varepsilon)-\varepsilon$ and
$\varepsilon(x_g)-x_g$ it follows:
\begin{equation*}
  \varepsilon(x_g)-x_g=\varepsilon(x_g-1)-(x_g-1)\
  \ \ \ \
  \varepsilon(1/2)-1/2=\varepsilon(-1/2)+1/2=\Delta_{1/2}
\end{equation*}
Equation~\ref{eq:290} can be finally expressed as:
\begin{equation}\label{eq:294}
\int_{-1/2}^{+1/2}\Gamma^p(x_g) x_g \mathrm{d}
x_g=\Delta_{1/2}=\varepsilon(-1/2)+1/2.
\end{equation}
$\Delta_{1/2}$ is just the missing part of the integration
constant of equation~\ref{eq:265a} that completes our initial
guess of $\varepsilon(-1/2)=-1/2$. To conclude, $\Gamma^p(x_g)$
and its mean value on a strip length allow the reconstruction of
asymmetric $\varphi(\varepsilon)$.

Excluding equation~\ref{eq:265c}, all the other equations of this
section are more general than the needs of
$\varphi(\varepsilon)$-reconstruction, and they will be used to
correct the systematic error due to the asymmetry in our noisy
case.

\subsection{Correction of the Systematic Error}

As we can see in figure~\ref{fig:figura60}, the systematic error
is the same for both algorithms. This is not unexpected. In
noiseless case, $x_{g2,3}=\pm 1/2$ is given by the same
$\varepsilon$-value. If the noise does not contribute in average
(and here this is the case for our acceptance of all the energy
values corrupted by the noise), the missing constant
$\varepsilon(x_{g2,3}=-1/2)+1/2$ is the same for $M_{\varepsilon
2}$ and $M_{\varepsilon 3}$.

A more precise strategy for the correction of $\varepsilon_{2,3}$
can be found studying their common forms. The final position
$\varepsilon_{2,3}(j)$ of the event $j$ is given by:
\begin{equation}
  \varepsilon_{2,3}(j)=x_{g2,3}(j)+\sum_{k=-L}^L\,\alpha_k(2,3)\mathrm{e}^{i2\pi k x_{g2,3}(j)}
\end{equation}
where $x_{g2,3}(j)$ is the COG of the event $j$ in their true
positions (more or less near to the maximum signal strip of the
event $j$). $\{\alpha_l(2,3)\}$ are the constants of the FS for
$\varepsilon_{2,3}$. We need the average of
$\varepsilon_{2,3}(j)-\varepsilon(j)$, the averages of
$\varepsilon_{2,3}(j)$ and $x_{g2,3}(j)$ are transparent. Due to
the non uniform distribution of $x_{gn}$ in the exponents, the
average of the FS is more complex. The $x_{gn}(j)$ are in the
global reference system, but the shift of the integer $\mu(j)$
does no modify the exponentials. The average over $\Gamma(x_{gn})$
due to the periodicity of the exponentials gives:
\begin{equation}
  \int_{-\infty}^{+\infty}\Gamma(x_{gn})\mathrm{e}^{i\,2\pi k x_{gn}}\mathrm{d}
  x_{gn}=\int_{-1/2}^{+1/2}\Gamma^p(x_{gn})\mathrm{e}^{i\,2\pi k x_{gn}}\mathrm{d}
  x_{gn},
\end{equation}
but for equations~\ref{eq:100} and ~\ref{eq:107} the average of
the FS becomes:
\begin{equation}
  \sum_{k=-L}^{+L}\alpha_k \int_{-1/2}^{+1/2}\Gamma^p(x_{gn})\mathrm{e}^{i\,2\pi k x_{gn}}\mathrm{d}
  x_{gn}=\alpha_0+\sum_{k=-L}^{+L}\alpha_k\alpha_{-k}i\,2\pi k.
\end{equation}
The bilinear term in $\alpha_k$ is zero due to the
$k$-antisymmetry of its elements, and the average
$M\varepsilon_{2,3}$ is:
\begin{equation}
  M\varepsilon_{
  2,3}=\alpha_0(2,3)+\sum_{j=1}^N\frac{x_{g2,3}(j)-\varepsilon(j)}{N}=
  \alpha_0(2,3)+M_{x2,3}^l+\sum_{j=1}^N\frac{\mu(j)-\varepsilon(j)}{N}
\end{equation}
Recalling the form~\ref{eq:265b} of $\alpha_0$ for any algorithm,
the integration on a triangle gives:
\begin{equation}
 \alpha_0=\int_{-1/2}^{+1/2}\mathrm{d} y[\int_{-1/2}^y\mathrm{d}
 y'\Gamma^p(y')-y]-\frac{1}{2}=
-\int_{-1/2}^{1/2}\mathrm{d} y' \Gamma^p(y') y'
\end{equation}
In the simulations (and in the data) $\alpha_l$, $M_{x2,3}^l$ and
$M\varepsilon_{ 2,3}$ are stochastic variables fluctuating around
their mean values, and the following equalities can experience
fluctuations due to the statistics:
\begin{equation}\label{eq:295}
\alpha_0(2,3)=-M_{x2,3}^l\ \ \ \ \Rightarrow\ \ \ M\varepsilon_{
2,3}=\sum_{j=1}^N\frac{\mu(j)-\varepsilon(j)}{N}
\end{equation}
Any reference to the used algorithm disappears in the expression
of $M\varepsilon_{ 2,3}$ and $M\varepsilon_{2}\approx
M\varepsilon_{3}$. The relation $\alpha_0(2,3)=-M_{x2,3}^l$ must
be handled with care. Different  probability distributions are
present in the two sides: $\alpha_0(2,3)$ is defined on a period
of the periodic $\Gamma^p(x_{g2,3})$,  $M_{x2,3}^l$ is defined on
$\Gamma(x_{g2,3})$. These differences are negligible in our
simulations of $x_{g2,3}$, but for $x_{g4}$ it is large and the
systematic error of $\varepsilon_4$ is different from that of
$\varepsilon_{2,3}$.

The correction of $M\varepsilon_{ 2,3}$ is obtained recalling
equation~\ref{eq:293}, that for an infinite sampling gives (for
asymmetric $p(x)$ the following integral is equal to the shift of
the COG of $p(x)$ respect to the strip center) :
\begin{equation}\label{eq:295a}
  \int_{-1/2}^{1/2}(x_g(\varepsilon)-\varepsilon)\mathrm{d} \varepsilon=0
  \ \Rightarrow \ \frac{1}{N}\sum_{j=1}^N
  x_{gn}(j)-\varepsilon(j)=0
\end{equation}
where $x_{gn}(j)$ approximate an infinite sampling containing all
the signal strips in the event. For $\theta\leq 20^\circ$,
$x_{g4}$ is negligibly different from an infinite sampling and
equation~\ref{eq:295a} can be recast:
\begin{equation}\label{eq:295b}
\sum_{j=1}^N\frac{x_{g4}(j)-\mu(j)}{N}+\sum_{j=1}^N\frac{\mu(j)-\varepsilon(j)}{N}=0
\ \ \ \ \rightarrow\ \ \ \
M_{x4}^l=-\sum_{j=1}^N\frac{\mu(j)-\varepsilon(j)}{N}.
\end{equation}
The required correction is the mean value of $x_{g4}(j)$ in the
reference system of the maximum signal strip we used to collect
all the events (in principle scattered in all the detector).

In figure~\ref{fig:figura60}, we see that even $M_{x3}^l$ can be a
good correction in many cases.

We pointed out a possible small non zero value of
$M_{x2,3}^l+\alpha_0(2,3)$, for this the complete correction
strategy is expressed by the following rule (for the algorithms
with two and three strips):

\begin{equation}\label{eq:297}
  \varepsilon_{2,3}^r(j)=\varepsilon_{2,3}(j)-M_{x2,3}^l-\alpha_0(2,3)+M_{x4}^l.
\end{equation}
For $\varepsilon_4(j)$ the correction reduces to the subtraction
of $\alpha_0(4)=-\Delta_{1/2}$ as imposed by equation~\ref{eq:294}
for the infinite sampling .

In figure~\ref{fig:figura70}, the dashed lines is obtained
shifting $\varepsilon_{2,3}(x_{g2,3})$ by
$-M_{x2,3}^l-\alpha_0(2,3)+M_{x4}^l$ and the new lines have a
better overlap with the noiseless $x_{g2,3}(\varepsilon)$.

\begin{figure}[h!]
\begin{center}
\includegraphics[scale=0.7]{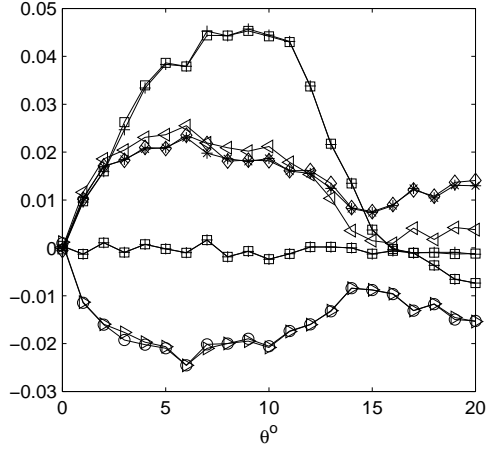}
\caption{\em Plots of mean values in function of $\theta$. Crosses
$M\varepsilon_{2}$, squares $M\varepsilon_{3}$, diamonds
$M_{x4}^l$, asterisks $M_{x3}^l$, triangles $M_{x2}^l$. The lines
around zero are $M\varepsilon_{2}^r$ and $M\varepsilon_{3}^r$.
}\label{fig:figura43}
\end{center}
\end{figure}

Equation~\ref{eq:297} is able to correct algorithms where
$\alpha_0(2,3)$ is widely different from $M_{x2,3}^l$ as in our
application of an aspect of the $\eta$-algorithm. In the
$\eta$-algorithm, the $\eta$-values are $0\leq\eta \leq 1$, and
the integration on $\Gamma(\eta)$ start from zero. For symmetric
$\varphi(x)$, zero is the right integration constant of
equation~\ref{eq:265a}, we use this zero constant for
$\varepsilon_{2,3}$ at all $\theta\leq 20^\circ$. In
figure~\ref{fig:figura43}, the systematic error of
$M\varepsilon_{2,3}$ is reported.

Even here $M\varepsilon_{2}(\theta)$ and
$M\varepsilon_{3}(\theta)$ are overlapping. An explanation of this
can be obtained observing that $x_{g2}=0$ and $x_{g3}=0$ are
obtained (in average) for the same $\varepsilon$-value, the
constant required to correct the systematic error. The differences
of $\alpha_0(2,3)$ from $-M_{x2,3}^l$ can be seen by the lowest
lines of figure~\ref{fig:figura43}, these are
$M\varepsilon_{2,3}-\alpha_0(2,3)-M_{x2,3}^l$ and are widely
different from $M\varepsilon_{2,3}$. In this case, the systematic
error is larger than the case with the integration starting at
$-1/2$.

The overlap of $M\varepsilon_{2}(\theta)$ with
$M\varepsilon_{3}(\theta)$ is a special property of the points
$\varepsilon(-1/2)$ and of $\varepsilon(0)$, it disappears with a
different stating point of the $\Gamma^p(x_g)$-integration.

\subsection{SD for the Floating Strip Side}
\begin{figure}[h!]
\begin{center}
\includegraphics[scale=0.7]{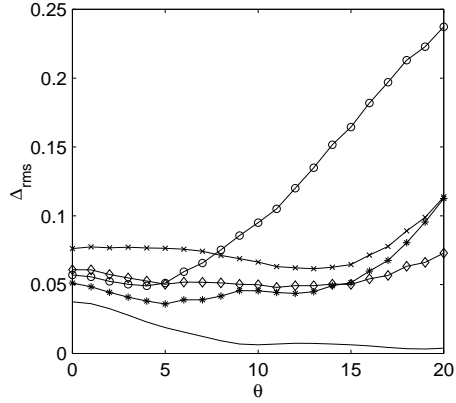}
\caption{\em Trends of the rms errors for various reconstruction
algorithms with $\theta^\circ$: crosses $x_{g3}-\varepsilon$,
circles $x_{g2}-\varepsilon$, diamonds
$\varepsilon_3-\varepsilon$, asterisks
$\varepsilon_2-\varepsilon$, continuous infinite sampling
analytical calculation }\label{fig:figura39}
\end{center}
\end{figure}
For $\theta=0^\circ$ the SD's of the various reconstructions are
smaller than that of normal strips. At increasing $\theta$, the
error is almost constant showing that the noise is the dominant
factor in reducing the resolution. The SD of $x_{g2}-\varepsilon$
starts to increase for $\theta>5^\circ$, the loss of information
creates a large gap in the $x_{g2}$-values around $x_{g2}=0$. The
reconstruction $\varepsilon_2$ is able to keep the error low, but
after $15^\circ$ the SD of $\varepsilon_3-\varepsilon$ is the
lowest. The infinite sampling noiseless $x_g$-error curve (of
equation~\ref{eq:262}) is always lower than all the other, as we
saw in the case of normal strips. The SD of $\varepsilon_{2,3}$
and $x_{g3}$ show an almost saturating trend up to $15^\circ$ and
after they start to increase.

\subsection{ Distributions of $\varepsilon_{2,3}-\varepsilon$}
As for the normal strip sensors, the SD (or the rms-error) is not
the best parameter to describe the distributions of the
differences $\varepsilon_{2,3}-\varepsilon$. Even for the floating
strip sensor, the distributions of the difference
$\varepsilon_{2,3}-\varepsilon$ have some similarities with a
Cauchy distribution. The histograms of $x_{g2,3}-\varepsilon$ are
similar to those of $\varepsilon_{2,3}-\varepsilon$. This is due
to the response function that approximates an ideal detector for
signal distributions of this size.

At increasing $\theta$ a systematic shift of the reconstructed
$x_{g2,3}(\varepsilon_{2,3})$ respect to the noiseless ones is
present in figure~\ref{fig:figura76}. A shift is evident even in
the histograms of the differences $\varepsilon_{2,3}-\varepsilon$.

\begin{figure}[h!]
\begin{center}
\includegraphics[scale=0.65]{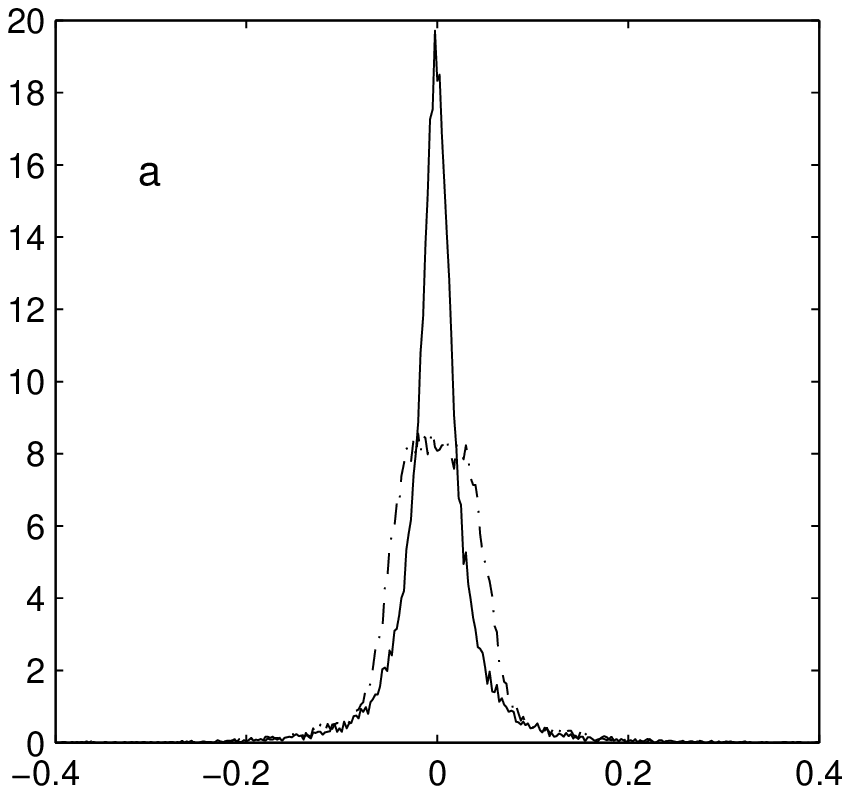}
\includegraphics[scale=0.65]{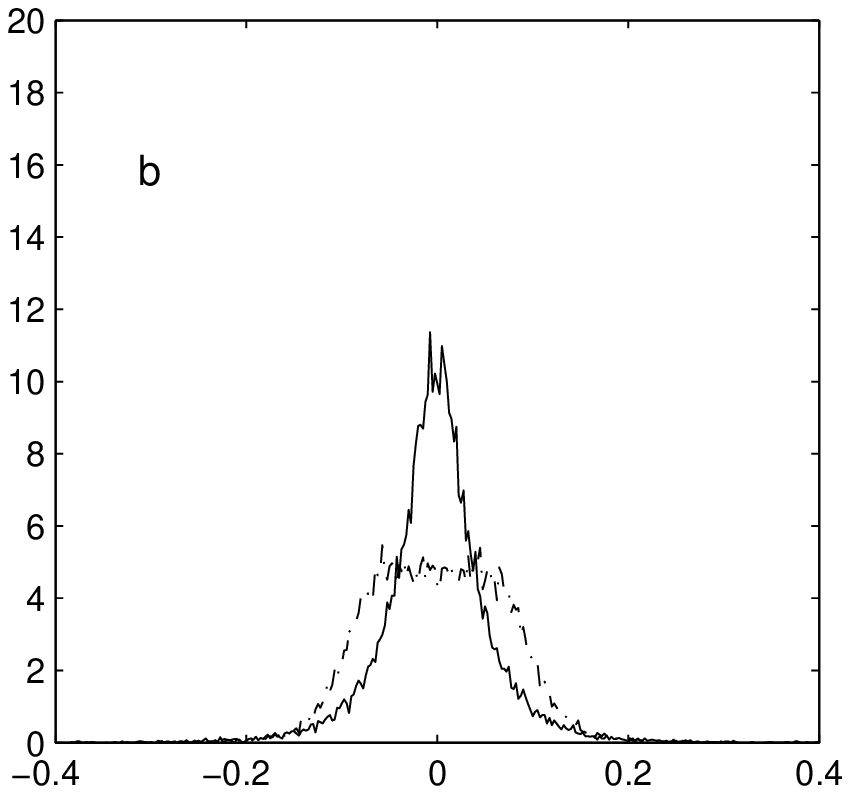}
\includegraphics[scale=0.65]{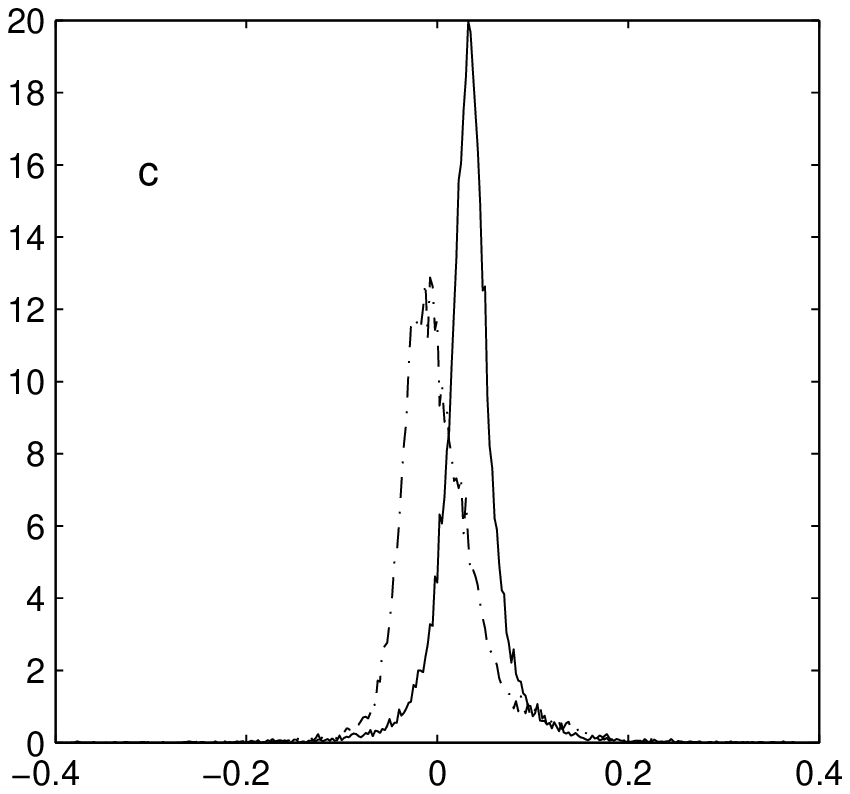}
\includegraphics[scale=0.65]{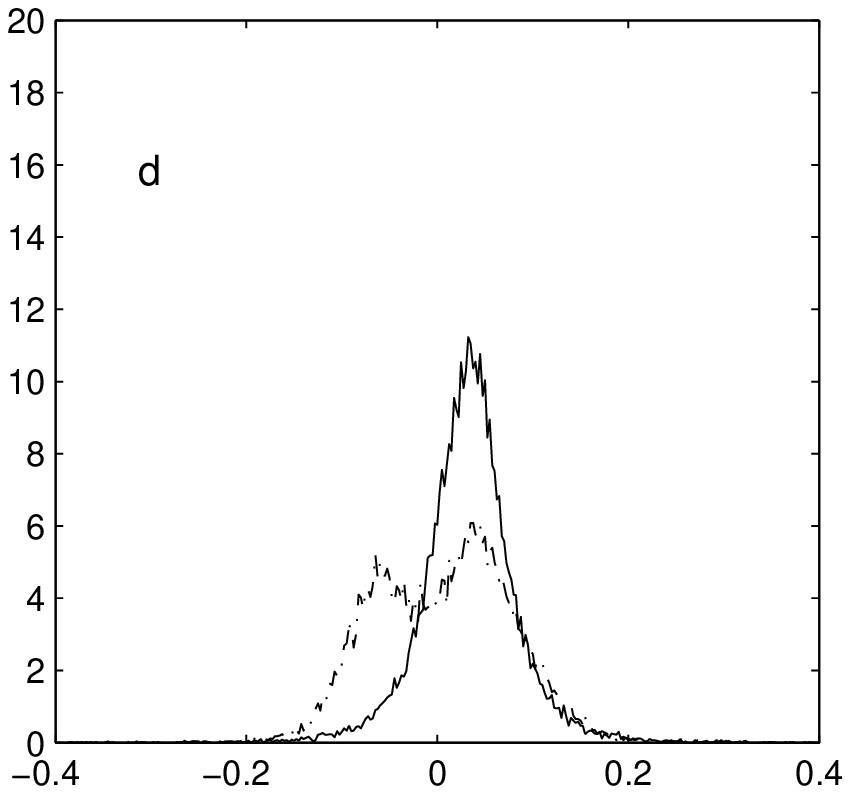}
\caption{\em To the left (a) it is the histogram of
$\varepsilon_2(x_{g2})-\varepsilon$ (continuous line) and of
$x_{g2}-\varepsilon$ (dash-dotted line) for normal strips and
$\theta=0^\circ$. To the right (b) the histogram of
$\varepsilon_3(x_{g3})-\varepsilon$ (continuous line) and of
$x_{g3}-\varepsilon$ (dash-dotted line). c) and d) are for
$\theta=3^\circ$. Here, a shift of the maximum respect to zero is
evident. } \label{fig:figura72}
\end{center}
\end{figure}

\begin{figure}[h!]
\begin{center}
\includegraphics[scale=0.6]{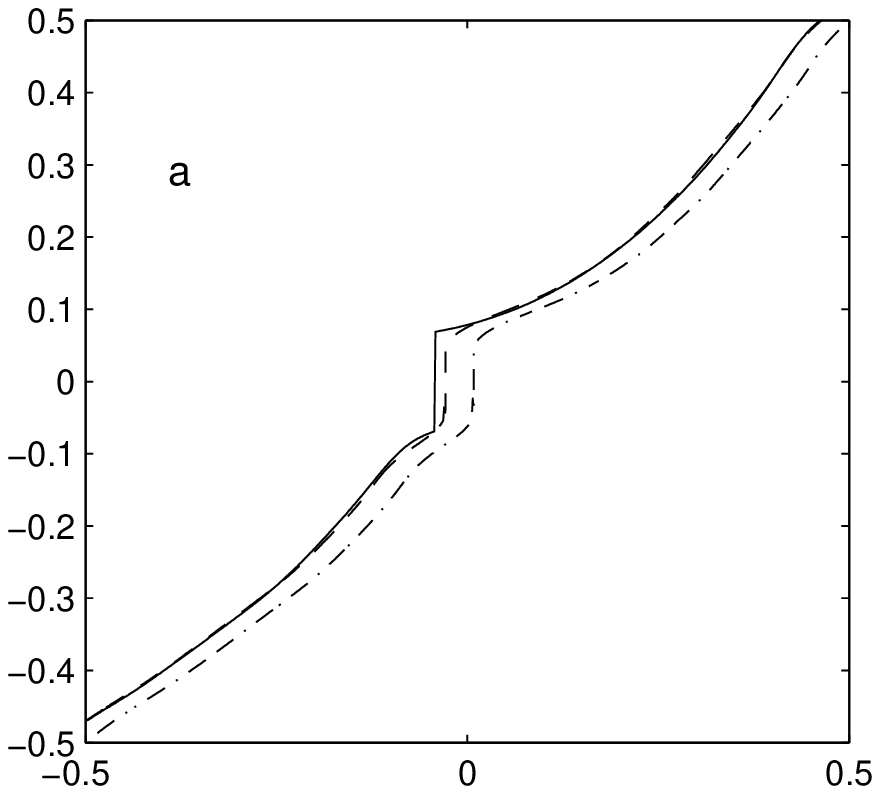}
\includegraphics[scale=0.6]{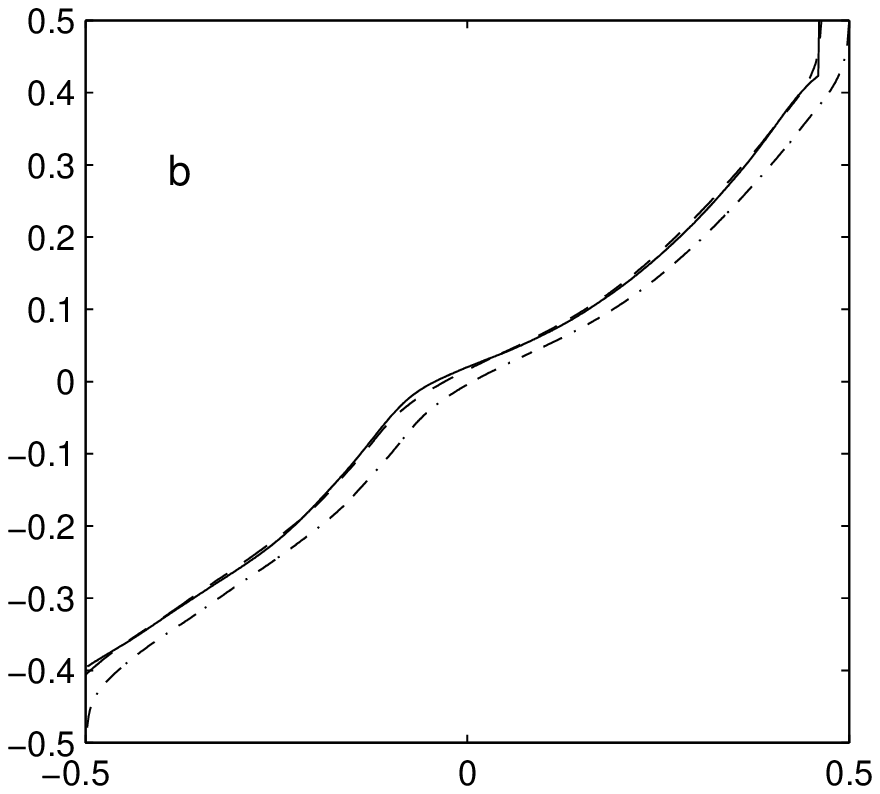}
\caption{\em To the left (a) is the plot of $x_{g2}$ vs
$\varepsilon_2(x_{g2})$ (dash-dot line) and $x_{g2}$ noiseless vs.
$\varepsilon$ (solid line) for $\theta=3^\circ$, the two curves do
not coincide as in figure~\ref{fig:figura21}, the dashed line is
$x_{g2}$ vs $\varepsilon_2^r(x_{g2})$ corrected as explained. This
curve has large overlaps with the noiseless one. To the right (b)
the plots of $x_{g3}$ versus $\varepsilon_3(x_{g3})$ (dash dot
line) and of $x_{g3}$ noiseless versus $\varepsilon$ (solid line)
and $x_{g3}$ vs $\varepsilon_3^r(x_{g3})$ (dashed line)}
\label{fig:figura76}
\end{center}
\end{figure}

At increasing $\theta$, figures~\ref{fig:figura72}c and d and
figure~\ref{fig:figura76} show shifts similar to those discussed
above for the normal strips, and their origin is identical. As
anticipated, their order of magnitude is that of the SD.

The histograms of $x_{g2,3}-\varepsilon$ do not show shifts of the
mean values, but small shifts are present. The mean value of
$x_g-\varepsilon$ is zero only for the infinite sampling case that
is well represented by $x_{g4}$ for $\theta\leq 20^\circ$. In the
algorithms with the suppression of signal strips the mean value of
$x_{gn}-\varepsilon$ has a non zero value that can be corrected
along the lines of equation~\ref{eq:295a} and~\ref{eq:295b}.

\begin{figure}[h!]
\begin{center}
\includegraphics[scale=0.6]{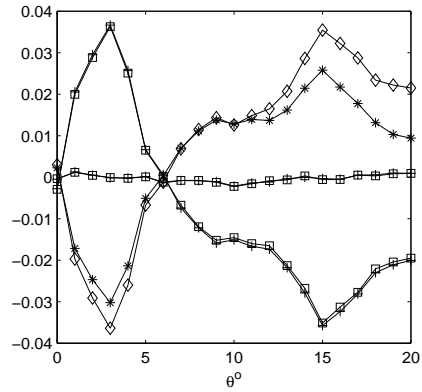}
\caption{\em Mean values in function of $\theta$, diamonds
$M_{x4}^l$, asterisks $M_{x2}^l$, squares $M\varepsilon_3$,
crosses $M\varepsilon_2$, the lines around zero are
$M\varepsilon_3^c$ e $M\varepsilon_2^c$. }\label{fig:figura59}
\end{center}
\end{figure}

The corrections of $M\varepsilon_{2,3}$ are in the form of
equation~\ref{eq:297}. In this sensor side we detected a small
signal loss for events around the floating strip, but the
corrections work well even here as it can be seen in
figure~\ref{fig:figura76}. Figure~\ref{fig:figura59} reports
$M\varepsilon_{2,3}(\theta)$ and $M_{x2,3,4}^l(\theta)$, the
overlap of $M\varepsilon_{2}(\theta)$ with
$M\varepsilon_{3}(\theta)$ indicates that an identical correction
is needed and  $M_{x4}^l$ is the correction. The approximate
identity $\alpha_0(2,3)=M_{x2,3}^l$ here is well verified.


\section{Conclusions}

The simulations we realized show useful applications of the
results of ref.~\cite{landi01}. It turns out sufficiently easy to
extract free parameters from the data. The response functions of
the strips are optimized for floating strip sensors and normal
sensors. For floating strip sensors and signal distributions of
this size, the obtained response function can be considered a
rough approximation of triangular function with base 2$\tau$. This
explains the better quality of the position reconstructions of
these detectors. The triangular response function (with base
2$\tau$) is the simplest response form of a detector (defined
ideal in ref~\cite{landi01}) where the COG of the hit strips is
the exact reconstruction algorithm. The simulations of non
orthogonal incoming directions (and asymmetric signals) have
immediately underlined the existence of a systematic error in the
standard applications of the $\eta$ algorithm. From the form of
our equations, the existence of this type of systematic error is
evident from the beginning. The simulations estimate its
amplitude, and they give some hint for its suppression. As in the
$\eta$ algorithm, the corrections are extracted from the data. A
fundamental step to demonstrate the form of the corrections is the
completion of a signal reconstruction theorem for an irregularly
sampled, non negative, asymmetric and duration limited signal.





\end{document}